\documentclass[aps,preprintnumbers,twocolumn,floatfix,nofootinbib]{revtex4}
\usepackage{epsfig}
\usepackage{amsmath}
\usepackage{graphicx,xcolor}
\usepackage{hyperref}
\newcommand{\sla}[1]{\slash\!\!\!{#1}}
\usepackage{ulem}

\newcommand{\T}[1]{\boldsymbol{#1}_{\text{T}}}

\newcommand{\Tsc}[2]{#1_{#2\text{T}}}
\newcommand{\Tscsq}[2]{#1^2_{#2\text{T}}}

\newcommand{\no}{\nonumber \\}
\newcommand{\parz}[1]{\ensuremath{\left(#1\right)}}
\newcommand{\order}[1]{\ensuremath{O\parz{#1}}}

\newcommand{\diff}[1]{\mathrm{d}#1}

\newcommand{\erefs}[2]{Eqs.~(\ref{e.#1})--(\ref{e.#2})}
\newcommand{\eref}[1]{Eq.~(\ref{e.#1})}
\newcommand{\fref}[1]{Fig.~\ref{f.#1}}

\newcommand{\sref}[1]{Sec.~\ref{s.#1}}
\newcommand{\aref}[1]{Appendix~\ref{a.#1}}

\begin{document}

\preprint{JLAB-THY-20-3185}

\title{Intrinsic Transverse Momentum and Evolution in Weighted Spin Asymmetries}

\author{Jian-Wei Qiu$^1$}
\thanks{Electronic address: jqiu@jlab.org \\
\href{https://orcid.org/0000-0002-7306-3307}{ORCID: 0000-0002-7306-3307}} 
\author{Ted~C.~Rogers$^{1,2}$}
\thanks{Electronic address: trogers@odu.edu \\ \href{https://orcid.org/0000-0002-0762-0275}{ORCID: 0000-0002-0762-0275}} 
\author{Bowen Wang$^3$}
\thanks{Electronic address: 0617626@zju.edu.cn  \\
\href{https://orcid.org/0000-0002-9732-6896}{ORCID: 0000-0002-9732-6896}} 

\affiliation{$^1$Jefferson Lab, 12000 Jefferson Avenue, Newport News, VA 23606, USA}
\affiliation{$^2$Department of Physics, Old Dominion University, Norfolk, VA 23529, USA}
\affiliation{$^3$Zhejiang Institute of Modern Physics, Department of Physics, Zhejiang University, Hangzhou, Zhejiang 310027, CHINA}

\begin{abstract}
The transverse momentum 
dependent (TMD) and collinear higher twist theoretical factorization frameworks are the most frequently used 
approaches to describing spin dependent hard cross sections weighted by and integrated over transverse momentum. Of particular interest is 
the contribution from small transverse momentum associated with the target bound state.
In phenomenological applications, this contribution is often investigated using transverse 
momentum weighted integrals that sharply regulate the large transverse momentum contribution, for example with Gaussian parametrizations. Since the result is a kind of hybrid of TMD and collinear (inclusive) treatments, it is important to establish if and how the formalisms are related in applications to weighted integral observables.
The suppression of a large transverse momentum tail, for example, can potentially affect the type of evolution that is applicable. 
We find that a naive version of a widely used identity relating the $k_T^2$-weighted and integrated Sivers TMD function to a 
renormalized twist-3 function has strongly ambiguous ultraviolet contributions, and that corrections to it are not necessarily perturbatively suppressed. 
We discuss the implications for applications, arguing in particular that the relevant evolution for transverse momentum weighted and integrated cross sections with sharp effective large transverse momentum cutoffs is of the TMD form rather than the standard renormalization group evolution of collinear correlation functions. 
\end{abstract}

\date{July 31, 2020}

\maketitle

\section{Introduction}
\label{s.intro}

Understanding fully the single transverse-spin asymmetries (SSA) of high energy scattering cross sections with the momentum transfer $Q\gg\Lambda_{\rm QCD}$ is still one of the most fascinating and challenging subjects in QCD since its discovery in hadronic $\Lambda^0$ production over 40 years ago \cite{Bunce:1976yb}.  The transverse SSA, defined as $A_N=(\sigma(S_T)-\sigma(-S_T))/(\sigma(S_T)+\sigma(-S_T))$, has been observed in many cross sections $\sigma(S_T)$, involving a single transverse hadronic spin $S_T$, and can be as large as 30-40\% in the forward region of hadronic single pion production \cite{Adams:1991rw,Bravar:1996ki,Aidala:2012mv}. This contradicted expectations about the size of the asymmetry that were based on early theoretical calculations \cite{Kane:1978nd}. With the parity and time-reversal invariance of QCD, it was recognized that the non-vanishing $A_N$ is a consequence of nonperturbative partonic motion and its correlation with the direction of the observed hadronic spin.  Thus, $A_N$ is a uniquely useful observable for probing a hadron's internal partonic structure and for studying quantum correlations between the partonic dynamics and emergent hadronic properties such as total spin \cite{Accardi:2012qut}.

\begin{figure}[h!]
  \includegraphics[width=0.4\textwidth]{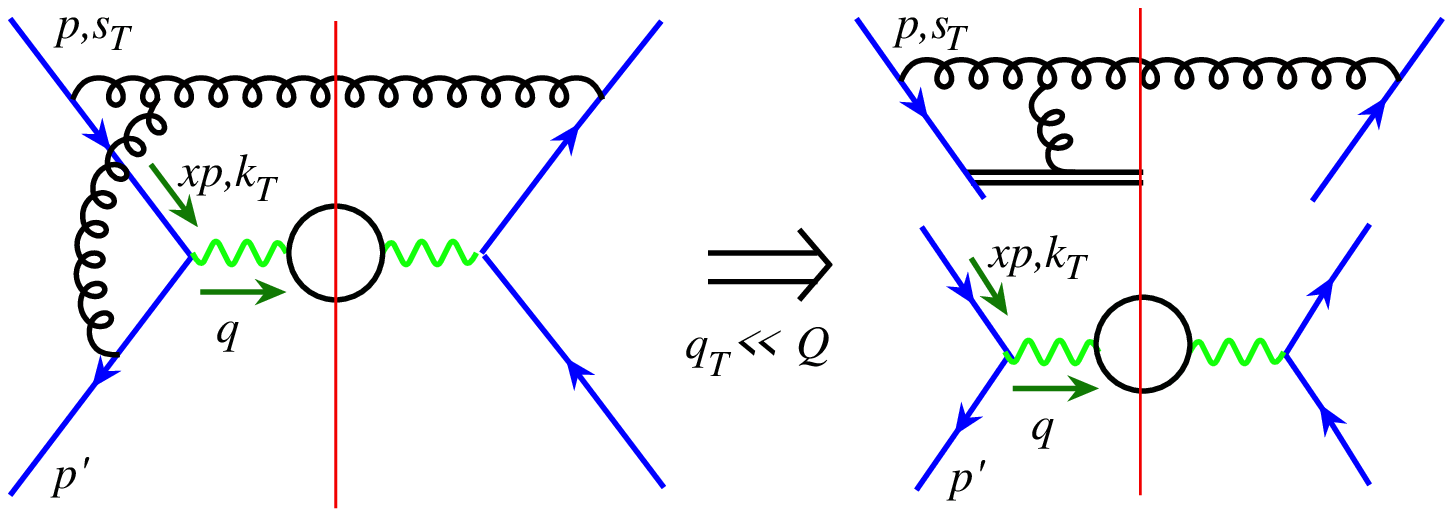}
  \\
  (a)
  \\
  \includegraphics[width=0.38\textwidth]{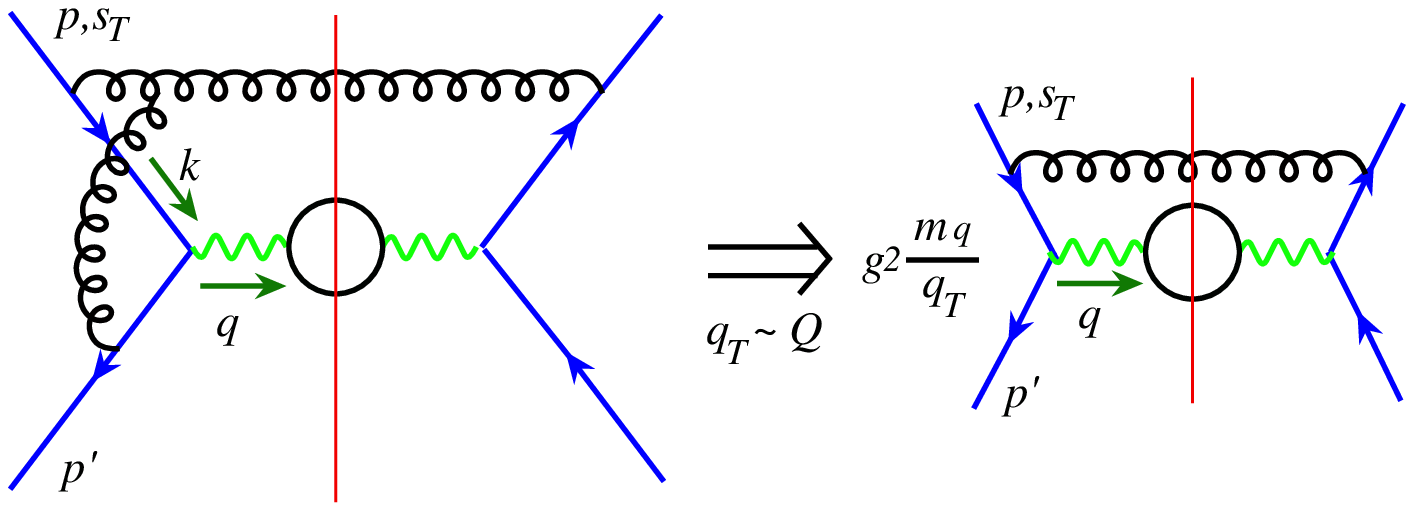}
  \\
  (b)
  \\
  \includegraphics[width=0.4\textwidth]{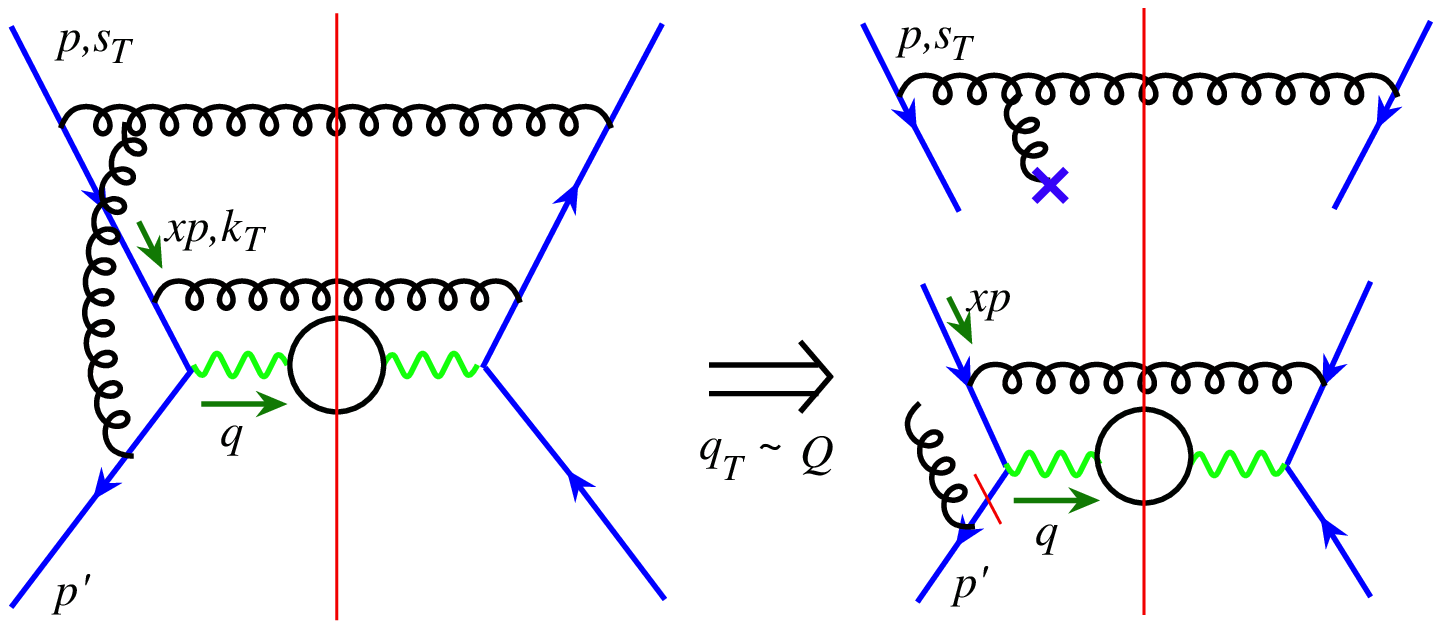} 
  \\
  (c)
\caption{Low order diagrams for $A_N$ of Drell-Yan lepton pair production by 
hard quark-antiquark annihilation: (a) TMD factorization when $k_T\sim Q_T \ll Q=\sqrt{q^2}$. 
(b) fixed-order perturbative QCD (pQCD) calculation when $Q_T \sim Q$. 
(c) Twist-3 collinear factorization when $k_T \ll Q_T\sim Q$.}
\label{fig:ssa4dy}
\end{figure}

The ability to understand $A_N$ in terms of the correlations between the partonic motion and hadronic spin relies on QCD factorization~\cite{Collins:1989gx} since any cross section with an identified hadron (and any corresponding asymmetry) is not perturbatively calculable in QCD.  A QCD factorization formalism for $A_N$ depends on the kinematics of measured cross section $\sigma(S_T)$.  As a typical two-scale observable, for example, Drell-Yan lepton pair production by 
hard quark-antiquark annihilation, as shown in Fig.~\ref{fig:ssa4dy} for partonic targets, can have very different factorization formalisms for  $\Delta\sigma(Q_T,Q;S_T)=\sigma(Q_T,Q;S_T)-\sigma(Q_T,Q;-S_T)$, defined as the difference between cross sections with the transverse spin flipped. In addition, it has two observed momentum scales associated with the virtual photon: its invariant mass $Q=\sqrt{q^2} \gg \Lambda_{\rm QCD}$ and the transverse momentum $Q_T$ with respect to the collision axis of $p$ and $p'$. When the active parton's transverse momentum is in the range of $k_T\sim Q_T\ll Q$, represented by Fig.~\ref{fig:ssa4dy}(a), intrinsic transverse momentum can be important and   
transverse momentum dependent (TMD) factorization is relevant. Then the contribution to $A_N$ from the diagram on the left can be approximately represented (schematically) by the TMD factorized expression on the right,
\begin{align}
&{}\Delta\sigma(Q_T,Q;S_T) \propto \nonumber \\ &{} \qquad \hat{\sigma}^{(0)}_{q\bar{q}\to l\bar{l}(Q)}
\otimes f_{\bar{q}/\bar{q}}^{(0),{\rm TMD}} \otimes k_T f^{\perp(1)}_{1T,q/q}
+\order{Q_T/Q} \, , \label{eq:tmddy}
\end{align}
where $\hat{\sigma}^{(0)}$ is the lowest order partonic Drell-Yan cross section, shown as the lower factorized diagram on the right of the arrow in Fig.~\ref{fig:ssa4dy}(a), 
$f_{\bar{q}/\bar{q}}^{(0), {\rm TMD}}(x',k'_T)$ is the zeroth order unpolarized TMD antiquark distribution of an antiquark, which is proportional to $\delta(1-x')\delta^2(k'_T)$ in lowest order perturbation theory,
$f^{\perp(1)}_{1T,q/q}(x,k_T)$ is the first order quark Sivers TMD function of a quark, given by the top diagram on the right of the arrow in Fig.~\ref{fig:ssa4dy}, and $\otimes$ indicates the convolution of the active parton's momentum, both longitudinal and transverse in this case.  The asymmetry, $A_N$, is generated by the non-vanishing Sivers function \cite{Sivers:1989cc,Sivers:1990fh,Brodsky:2002rv}.  

But, when $Q_T\sim Q$, the same diagram, now symbolized by Fig.~\ref{fig:ssa4dy}(b), would give a leading fixed order contribution to $A_N$ while the loop on the left generates the needed phase and the quark mass $m_q$ generates the spin flip for the $A_N$. This leads to an asymmetry proportional to 
$g_s^2 m_q/Q_T$ with strong coupling constant $g_s$, 
which was predicted to be very small in view of $m_q\ll Q_T\sim Q$ \cite{Kane:1978nd}.  

At the same $Q_T\sim Q$, however, additional mechanisms can generate transverse SSAs, and these are symbolized in Fig.~\ref{fig:ssa4dy}(c).  When the transverse momentum of the active parton for the hard scattering to produce the lepton pair is in the range $k_T\ll Q_T\sim Q$, the formally higher order diagram on the left can be factorized in terms of twist-3 collinear factorization
\begin{align}
&{} \Delta\sigma(Q_T,Q;S_T) \propto \nonumber \\ &{} \qquad \hat{H}^{(1)}_{q(g)\bar{q}\to l\bar{l}(q)}
\otimes f_{\bar{q}/\bar{q}}^{(0), {\rm coll}} \otimes T^{(1)}_{q(g)/q}
+\order{\Lambda_{\rm QCD}/Q} \, , \label{eq:tw3dy}
\end{align}
where $\hat{H}^{(1)}$ is the lowest order partonic hard part to produce the SSA of high-$Q_T$ lepton pair production. This is shown as the bottom diagram on the right of the arrow in Fig.~\ref{fig:ssa4dy}(c), with the unpinched pole of the antiquark-line having a (red) bar to indicate the needed phase. The active quark-gluon composite state allows for the helicity to flip between the left and the right of the cut in this diagram, even with zero quark mass. 
The $f_{\bar{q}/\bar{q}}^{(0), {\rm coll}}(x')$ is the zeroth order unpolarized twist-2 collinear antiquark distribution of an antiquark at lowest order of perturbation theory, which is proportional to $\delta(1-x')$,
$T^{(1)}_{q(g)/q}(x)$ is the first order twist-3 quark-gluon correlation function of a quark, given by the top diagram on the right, and $\otimes$ indicates the convolution of active parton's longitudinal momentum fractions.  The typical transverse momenta of active partons here, which are expected to be much smaller than the hard scale, $Q_T\sim Q$, are integrated into the twist-3 quark-gluon correlation function, whose size is determined by the imbalance of quark motion generated by the color Lorentz force (the gluon) in defining the twist-3 quark-gluon correlation functions \cite{Efremov:1984ip,Qiu:1991pp,Qiu:1991wg,Qiu:1998ia}.

Both TMD and twist-3 collinear factorization formalisms, in Eqs.~(\ref{eq:tmddy}) and (\ref{eq:tw3dy}) respectively, have been argued to be valid to all orders in QCD perturbation theory for their respective kinematical regimes \cite{Collins:1984kg,Collins:2011zzd,Ji:2004wu,Ji:2004xq,Qiu:1990xy,Qiu:1990cu}.  In an overlap region where $\Lambda_{\rm QCD}\ll Q_T \ll Q$, 
the TMD and twist-3 collinear factorization formalisms for the SSAs were shown to be consistent with each other \cite{Ji:2006ub,Ji:2006vf,Scimemi:2019gge} 
when the active parton $k_T$ and the phase of the Sivers TMD function are perturbatively generated by the twist-3 mechanism.

Both TMD and twist-3 collinear factorization approaches have also been used frequently to describe the transverse {\it moment} of two-scale spin dependent hard cross sections and their asymmetries, by integrating over transverse momentum $Q_T$ while weighting by a single power of ${\bf Q}_T$, leaving the observables with only a single large momentum transfer $Q$ \cite{Kang:2012ns,Dai:2014ala,Gamberg:2017jha,Xing:2019nof,Luo:2020hki}.  
In principle, the moments (or the asymmetries of the moments) of $Q_T$-distributions should be described by a QCD collinear factorization formalism, if one exists, since the active parton's $k_T$ should be much less than the single hard momentum transfer $Q$.  In practice, however, both factorization approaches have been adopted for evaluating the moments of the $Q_T$-distributions.  For example, an unpolarized Drell-Yan-like cross section $\frac{d\sigma}{dQ^2}$ is often calculated in terms of QCD collinear factorization with perturbatively calculated hard parts $\frac{d\hat{\sigma}_{ij}}{dQ^2}$ convoluted with two twist-2 collinear parton distribution functions (PDFs) $f_{i/H}(x)$~\cite{Collins:1989gx}. (Here we suppress the factorization scale and active parton flavor indexes, $i,j=q,\bar{q},g$.) The same observable can be viewed as the $0^{\rm th}$ moment of the $Q_T$-distribution, $\frac{d\sigma}{dQ^2}=\int dQ_T^2\, (Q_T^2)^0 \frac{d\sigma}{dQ^2 dQ_T^2}$, with the $\frac{d\sigma}{dQ^2 dQ_T^2}$ evaluated in terms of the TMD factorization formalism and unpolarized TMD pdfs $f_{i/H}(x,k_T)$ when $k_T\sim Q_T \ll Q$, along with a proper matching when $Q_T$ becomes larger ($Q_T\sim Q$) to a cross section calculation $\frac{d\sigma^{\rm Pert}}{dQ^2 dQ_T^2}$ performed in terms of QCD collinear factorization with twist-2 collinear PDFs~\cite{Collins:1984kg}.  Both approaches are well-defined within QCD perturbation theory and within the frameworks of their corresponding factorization theorems.

Of course, the above remarks apply similarly to other processes with a transversely polarized hadronic target, particularly SIDIS with its typically smaller $Q$ and higher sensitivity to nonperturbative hadronic structure.

A commonly used relation between TMD pdfs and twist-2 collinear PDFs,
\begin{equation}
\int \diff{^2\T{k}{}}{} f_{i/H}(x,k_T)  = f_{i/H}(x) \, , 
\label{e.closure}
\end{equation}
 connects the two approaches to each other, up to  $\order{\alpha_s}$-suppressed terms associated with  different ways of including high order corrections \cite{Berger:2003pd}.  When the full TMD factorization formalism is used for the region of $Q_T \ll  Q$, and optimized for the region $\Lambda_{\rm QCD} \ll Q_T \ll  Q$ with resummed $\ln(Q^2/Q_T^2)$-enhanced effects taken into account, the cross section as a $0^{\rm th}$ moment receives corrections to \eref{closure}, as demonstrated for inclusive Higgs production in a Drell-Yan-like process \cite{Berger:2003pd}.  

It has been proposed that the TMD and twist-3 collinear factorization approaches to describing the transverse moment of the two-scale spin dependent hard cross sections and their asymmetries are connected through a well-known relation between the Sivers TMD function $f_{1T,q/H}^{\perp}(x,\Tsc{k}{})$ of hadron $H$ and the twist-3 quark-gluon correlation function $T_{q(g)/H}(x)$ \cite{Boer:2003cm},
\begin{equation}
\int \diff{^2\T{k}}{} \frac{\Tscsq{k}{}}{M^2} f_{1T,q/H}^{\perp}(x,\Tsc{k}{}) =  - \frac{1}{M} T_{q(g)/H}(x) 
\, ,
\label{e.basic}
\end{equation}
in an analog to the relation in Eq.~(\ref{e.closure}), 
where factors of the hadron mass (labeled $M$) are included by convention to make both sides dimensionless. To simplify notation, we have dropped the usual second argument of the twist-3 quark-gluon correlation function $T_{q(g)/H}(x,x)$ since for our purposes we will only be interested in the case where both active quark momentum fractions are equal. 
For the relative minus sign in~\eref{basic}, 
the Wilson line in the Sivers TMD function should be understood to point in the direction relevant to lepton-hadron semi-inclusive deep inelastic scattering (SIDIS) \cite{Kang:2011hk}, which would require an extra minus sign in Eq.~(\ref{e.basic}) if we prefer to use the Sivers TMD function extracted from the the Drell-Yan type processes.  
In this paper, we try to verify the relation in Eq.~(\ref{e.basic}), and to understand how it is  similar or different from the unpolarized analog in Eq.~(\ref{e.closure}).  

The moment of the Sivers function on the left side of \eref{basic} arises naturally in studies of the moment or weighted transverse SSAs.  For example, the TMD factorized expression in Eq.~(\ref{eq:tmddy}) can be used to evaluate the ${\bf Q}_T$-weighted asymmetry if one assumes it is approximately valid for the full range of ${\bf Q}_T$-integration, that is, if one neglects the $Q_T \sim Q$ ``Y-term'' correction and assumes exact validity for \eref{basic}. This results in a factorized expression proportional to the integral on the left side of \eref{basic} \cite{Boer:2003cm}.

The equality in \eref{basic} is widely understood to imply that 
$T_{q(g)/H}(x)$ and $f_{1T,q/H}^{\perp}(x,\Tsc{k}{})$ 
are essentially different ways 
of representing similar physics~\cite{Kang:2012xf,Metz:2014bba,Gamberg:2017gle}, namely that of intrinsic non-perturbative parton transverse momentum inside a hadron target around $k_T \sim \Lambda_{\rm QCD}$. 
This view
has motivated various interpretations
of experimental data, 
including, for example, suggestions of tension in the phenomenology of the Sivers effect~\cite{Kang:2011hk,Kang:2012xf}. 
Equation~\eqref{e.basic} is also a common ingredient in phenomenological applications of twist-3 factorization because practical functional 
representations of 
the twist-3 quark-gluon correlation function 
are obtained via \eref{basic} from phenomenological extractions of the Sivers function~\cite{Kanazawa:2014dca}. 
It has also been suggested that~\eref{basic} provides a kind of loophole around the problems with 
TMD factorization that arise in certain processes~\cite{Gamberg:2010tj}.

In \eref{basic}, both the Sivers TMD function $f_{1T,i/H}^{\perp}(x,\Tsc{k}{})$ and the twist-3 correlation function $T_{q(g)/H}(x)$ are non-perturbative but could in principle be extracted from physically measured SSAs.  If the $\diff{^2\T{k}}{}$-integration of a measured $f_{1T,i/H}^{\perp}(x,\Tsc{k}{})$ weighted by $\Tscsq{k}{}$ converges, then the relation in \eref{basic} can be tested for its $Q^2$ dependence as well as its $x$ dependence.  However, the relation in \eref{basic} is often used in the literature as an identity to replace one side by the other side to help in the extraction of the Sivers TMD functions (or twist-3 correlation functions), and thus does not treat them as two different functions. Therefore, the precise reliability of the relation in \eref{basic} can impact on-going community efforts to extract non-perturbative TMD correlation functions and  to explore hadron's internal partonic structure and its correlation to the emergent hadronic properties.

In phenomenological applications, an ambiguity immediately arises as to what type of 
$Q^2$-dependence or scale evolution 
should be expected for 
the weighted integral on the left side of \eref{basic} \cite{Alexeev:2018zvl}. 
Taken literally, the right side of the equation implies that the $Q^2$-dependence 
should follow from a DGLAP-type
evolution of twist-3 quark-gluon correlation functions~\cite{Kang:2008ey,Braun:2009mi} 
since $T_{q(g)/H}(x)$ should be extracted from the observed $A_N$ factorized in terms of
the twist-3 collinear factorization.
By contrast, 
the $f_{1T,i/H}^{\perp}(x,\Tsc{k}{})$ is to be extracted from the observed 
$A_N$ differential in transverse momentum and factorized in terms of TMD factorization, whose $Q^2$-dependence 
should follow the Collins-Soper style of evolution \cite{Collins:1981va,Collins:1984kg}, and without a full treatment of the large ${\bf Q}_T \sim Q$ tail
the additional transverse momentum integral 
would not change this $Q^2$-dependence to the DGLAP-type. 

Like all QCD factorization formalisms, both the TMD and twist-3 collinear factorization theorems for SSAs are 
constructed such that collinear and infrared (IR) sensitivity is automatically removed from the partonic scattering process and placed in the non-perturbative long-distance but universal TMD functions and twist-3 quark-gluon correlation functions respectively.
The predictive power of the TMD and twist-3 collinear factorization in Eqs.~(\ref{eq:tmddy}) and (\ref{eq:tw3dy}) relies on: (a) the universality of the Sivers TMD functions and twist-3 collinear quark-gluon correlation functions and, by extension, (b) their abilities to systematically remove the collinear and infrared sensitivities of the corresponding partonic scattering to ensure the infrared safety of $\hat{\sigma}$ in Eq.~(\ref{eq:tmddy}) and $\hat{H}$ in Eq.~(\ref{eq:tw3dy}) order-by-order in QCD perturbation theory at all applicable momentum scales. 
Given the difference in operators defining the Sivers TMD function and the twist-3 quark-gluon correlation function, it is not immediately 
clear that one should expect $T_{q(g)/H}(x)$ and $f_{1T,q/H}^{\perp}(x,\Tsc{k}{})$ to have comparable 
non-perturbative small transverse momentum behavior, 
since the partonic versions of such objects and their scale evolution are clearly qualitatively different beyond the tree-level \cite{Aybat:2011ge,Kang:2008ey}.  The question is whether a weighted $k_T$-integration of $f_{1T,q/H}^{\perp}(x,\Tsc{k}{})$ like \eref{basic} would make them to be the same.  

Furthermore, in order to apply QCD factorization to the moment of spin dependent hard cross sections and their asymmetries beyond the tree-level in perturbative calculations, the operators that define Sivers TMD functions and twist-3 correlation functions in \eref{basic} should be the renormalized ones, and the renormalization of corresponding non-local operators needs to be specified.
Otherwise, the derivation of \eref{basic} involves manipulations with infinite quantities~\cite{Boer:2003cm}. So, 
in view of the widespread use of \eref{basic} it is important to characterize possible violations to it that might become apparent 
once the divergent behavior is taken into account. 
Indeed, the violation of \eref{basic} as an exact statement is already
well-known (see, for example, Ref.~\cite{Kang:2010hg}, along with the discussion there regarding sensitivity 
to large $k_T$ cutoff schemes). 
In particular, the removal of ultraviolet (UV) regulators does not generally commute with the evaluation of transverse momentum 
integrals. 
However, it is typically assumed that, after $k_T$-cutoffs are in place, 
violations to \eref{basic} correspond to small perturbative corrections and that it can be viewed as a kind of zeroth order approximation.

There are a number of open questions in the treatment of 
factorization for weighted inclusive observables generally, and we 
do not intend to address them all here. Indeed, with only one large momentum scale observed, a complete derivation of collinear factorization for fully inclusive weighted moments in terms of twist-3 functions alone does not yet, to our knowledge, exist.  Instead, we will 
highlight particular issues that arise by focusing 
on the properties of individual parton correlation functions when their definitions are taken literally. Nevertheless, we emphasize that, for implementations that focus on the small or nonperturbative transverse momentum region while suppressing the large transverse momentum tail, factorization with TMD correlation functions is natural.  

Within the assumption that all operator matrix elements are calculated using standard renormalization, 
we will argue using an explicit calculation that the breakdown of \eref{basic} is not perturbatively suppressed in the normal sense,
and is sensitive instead to a collinear regulator.  
We propose, therefore, to take \eref{basic} as a definition for the UV behavior of $T_{q(g)/H}(x)$ rather than as a derived result, at least for those observables that focus on the small transverse momentum region. Moreover, if transverse momentum cutoffs are sharp enough to retain sensitivity to non-perturbative intrinsic transverse momentum, as with, for example, narrow Gaussian parametrizations, then evolution of the corresponding weighted and integrated asymmetries should be for TMD functions rather than through collinear evolution.  
The Gaussian (or similar) ansatz approach to TMD phenomenology 
has met with significant success in applications~\cite{Anselmino:2012aa,Boglione:2018dqd,Cammarota:2020qcw}, and is an approach that maintains a more natural link to intrinsic nonperturbative physics than those that focus more on accurately describing a broad perturbative transverse momentum tail. 

Similar identities to \eref{basic} are used to relate other kinds of twist-3 collinear and TMD functions, for example the Collins fragmentation function~\cite{Metz:2012ct,Yuan:2009dw,Kang:2015msa},
and there are many similar proposed relations between twist-3 and TMD correlation functions (e.g., Eqs.(C13-C15) of~\cite{Mulders:1995dh}).  Thus, our results potentially impact the study of weighted-integrated correlation functions more broadly. 

The rest of this paper is organized as follows:  In the next section, we introduce our conventions for the renormalization of parton distribution functions (PDFs) and, in general, parton correlation functions.  As an example, and to set up later discussions of \eref{basic}, in Sec.~\ref{comparison} we further 
discuss the relation in~\eref{closure} relating 
spin averaged TMD PDFs and collinear PDFs.
In Sec.~\ref{s.setup}, we specify how the proposed identity in \eref{basic} is to be tested, and show the violation of the identity in terms of an explicit lowest order calculation in perturbative QCD in Sec.~\ref{s.direct}.  Finally, we discuss our results and our proposal for the treatment of the evolution of weighted asymmetries in Sec.~\ref{s.discussion}.

\section{Renormalization}
\label{renorm}

We will refer to the renormalization of PDFs in the standard sense of a renormalization of a generalized  operator product. 
So, for example, the renormalized collinear PDF for a quark in a hadron is 
\begin{equation}
\label{e.pdfren}
f_{i/H}(x;\mu) = Z_{ij} \otimes f_{j/H,0} \, ,
\end{equation}
where $i,j$ represent the quark flavor.
The bare PDF $f_{j/H,0}(x)$ has the usual definition of a PDF, but defined with bare fields.
The $\otimes$ denotes the usual convolution products 
over longitudinal momentum fractions that appear in collinear factorization, and $\mu$ is the usual renormalization group scale. Our calculations that use dimensional regularization will work in  dimension $D = 4 - 2 \epsilon$
and use a generalized minimal subtraction scheme for renormalization, in which case the $Z_{ij}$ beyond zeroth 
order consist only of $\epsilon$-poles with mass-independent coefficients. 

It is important to note that for higher twist operators renormalization 
can mix with operators of lower dimension.

Renormalization is not the only valid way to define the ultraviolet behavior of collinear correlation functions, but it comes with many desirable features, including the automatic cancellation of lightcone divergences and order-by-order validity of number and momentum sum rules. We therefore view it as the default approach. 

Renormalization works similarly for TMD PDFs, 
though an extra kind of generalized renormalization is needed in association with lightcone divergences \cite{Collins:2011zzd}. Such issues will not arise directly in this paper, however. 

For the message of this paper to be clear, it is important to 
recall that with the renormalization approach to PDFs, virtual and real UV 
divergences need to be consistently regulated in the same way -- see \sref{discussion} below for more on this. 

\section{Comparison with unpolarized case}
\label{comparison}

The equality in Eq.~(\ref{e.closure}) relating unpolarized TMD pdfs and collinear PDFs is similar to the relation in Eq.~(\ref{e.basic}) in the sense that a moment of the TMD pdf is related to a corresponding collinear PDF. But the two equalities in Eqs.~(\ref{e.closure}) and (\ref{e.basic}) are also fundamentally different 
in the nature of the operators involved.  

For the relation in Eq.~(\ref{e.closure}), 
the non-local operators defining the TMD pdfs on the left-hand-side and 
the non-local operators of corresponding collinear PDFs on the right-hand-side 
share the same leading twist local operators when the operator product expansion (OPE) 
is applied to these non-local operators.  That is, both sides of the relation in Eq.~(\ref{e.closure}) 
share the same leading twist, leading order perturbative collinear and UV behavior.  
As discussed earlier, the same collinear sensitivities in perturbative calculations from the two sides of the equality 
is a requirement for factorization if both sides 
of \eref{closure} are to be equally valid definitions for the collinear PDF. 
The integration over $k_T^2$, specifically the transverse momentum flow 
between the active quark in the scattering amplitude and its complex conjugate, 
picks up the leading twist operators with no transverse separation, 
which are logarithmically UV divergent and require renormalization. Consequently, 
differences between the two sides of \eref{closure} could potentially include the effect of different choices (or schemes) for renormalizing the perturbative leading twist UV divergence, and how this differs from the UV regulator of the integration over the
active parton's transverse momentum, $k_T^2$.  
This leading twist scheme dependence does not change the 
collinear sensitivities of either side in Eq.~(\ref{e.closure}).  That is, any possible
difference between the two sides of the relation in Eq.~(\ref{e.closure}) 
is infrared insensitive or perturbatively calculable.    
Before turning to the extra complications that might arise with \eref{basic}, 
we quantify the relation in Eq.~(\ref{e.closure}) in the rest of this section. 

As is well-known,~\eref{closure} 
is actually valid up to perturbative corrections for renormalized PDFs (both collinear and TMD) when a cutoff $k_c$ is imposed on the transverse momentum integral and if TMD pdfs are defined in any of the usual senses~\cite{Collins:1981uw,Collins:2011zzd,Diehl:2015uka,Aybat:2011ge,Rogers:2015sqa,Stewart:2009yx,Becher:2010tm, Becher:2011xn, Becher:2012yn, GarciaEchevarria:2011rb, Echevarria:2012js, Echevarria:2014rua, Chiu:2012ir,Li:2016axz} that are currently used. We may state this explicitly by first defining 
\begin{equation}
\Delta f(k_c) 
\equiv \pi \int_0^{k_c^2} \diff{\Tscsq{k}{}}{} \,   
f_{i/H}(x,\Tsc{k}{};\mu) - f_{i/H}(x;\mu) \, , 
\label{e.deltdef}
\end{equation}
where the definition of $f_{i/H}(x,\Tsc{k}{};\mu)$ is any of the standard TMD definitions, 
and $f_{i/H}(x;\mu)$ is 
the standard renormalized parton density. Then it is straightforward to verify that the following factorization holds: 
\begin{align}
& \Delta f(k_c) = \nonumber \\
&\; \sum_{ij} \mathcal{C}_{ij}\parz{x/x',\alpha_s(\mu),L} \otimes f_{j/H}(x;\mu) + \order{\frac{\Lambda^2_{\rm QCD}}{k_c^2}} \,  ,\label{e.yuk}
\end{align}
where the $\mathcal{C}_{ij}$ are mass-independent generalized functions that depend on 
$\mu$ only through $\alpha_s(\mu)$ and powers of the logarithm
\begin{equation}
L(k_c/\mu) \equiv \ln \parz{\frac{\mu^2}{k_c^2}} \, . \label{e.logdef}
\end{equation}
The $\mathcal{C}_{ij}$ start at order $\alpha_s(\mu)$ or higher. 
Therefore, as long as the cutoff $k_c$ is fixed roughly at order $\mu$, corrections to \eref{closure} are suppressed by 
at least a power of $\alpha_s(\mu)$. When both the $\alpha_s(k_c)$-suppressed and $(\Lambda^2_{\rm QCD}/k_c^2)$-suppressed 
terms in \eref{yuk} are dropped, the identity in \eref{closure} is restored. Verifying the 
above is possible to do directly in renormalizable model field theories or in pQCD order-by-order. 

The purpose of the discussion above is to make statements about relations
like \eref{closure} or \eref{basic} holding at ``lowest order" precise.  
Then in the next two sections we explain why a statement analogous to \eref{yuk} fails for \eref{basic} if applied to ordinary renormalized correlation functions.

\section{Spin Dependent Case}
\label{s.setup}
Now we return to \eref{basic}. 
The general form of the operator definition of the (pole part of the) twist-3 quark-gluon correlation function is
\begin{align}
&{} T_{i(g)/H}(x) = g_s\, \epsilon^{S_T \alpha}g_{\alpha \beta} \nonumber \\
&{} \times  \int \frac{\diff{\xi^-}{} \diff{\eta^-}{} }{4 \pi}
e^{i x P^+  \xi^-} \langle P,S| \bar{\psi}_i(0) G^{\beta +}(\eta^-) \gamma^+ \psi_i(\xi^-) |P,S\rangle \, . \label{e.qs}
\end{align}
$S_T$ is the transverse spin of the target and $G^{\mu \nu}$ is the gluonic field strength tensor. The analog of \eref{pdfren} is a 
renormalized twist-3 quark-gluon correlation function:
\begin{align}
\label{e.qsren}
T_{i(g)/H}(x;\mu) 
=& 
\sum_{ij} Z_{a,ij} \otimes T_{j(g)/H,0} 
\nonumber \\
+& 
\sum_{b,ij} m_{b,ij} Z_{b,ij} \otimes h_{j/H,0,b}\, .
\end{align}
The $T_{i(g)/H,0}$ are defined as in \eref{qs}, but here specifically with bare fields,  
the $h_{i/H,0,b}$ are any of the possible 
lower 
twist bare collinear operator
matrix elements that might be necessary in the renormalization, and 
the $m_{b,ij}$ are the renormalized masses of any of the fields. As before, $i$ and $j$ are parton 
flavor indices. The $Z_a$ and $Z_b$ coefficients are renormalization factors respectively for the bare collinear twist-3 function $T_{j(g)/H,0}$ and any other lower dimension operators. In dimensional regularization with generalized minimal subtraction, they are only mass-independent poles in $\epsilon$. 
  
The analog of \eref{deltdef} for \eref{basic} is 
\begin{align}
&{}\Delta f_{1T}^\perp(k_c) \nonumber \\
&{} \equiv \pi \int_0^{k_c^2} \diff{\Tscsq{k}{}}{} \frac{\Tscsq{k}{}}{M^2} f_{1T,i/H}^{\perp}(x,\Tsc{k}{};\mu) 
 + \frac{1}{M} T_{i(g)/H}(x;\mu) \, . \label{e.deltdef2}
\end{align}  
If a version of \eref{basic} held at zeroth order, then it would have to be possible to 
express $\Delta f_{1T}^\perp(k_c)$ in 
the following factorized way
\begin{align}
&{} M \Delta f_{1T}^\perp(k_c) \stackrel{??}{=} 
\sum_{ij} \mathcal{C}_{ij}\parz{x/x',L,\alpha_s(\mu)} \otimes T_{j(g)/H}(x;\mu)  \nonumber \\
&{} \qquad + \sum_{b,ij} m_{b,ij} \mathcal{C}_{b,ij}\parz{x/x',L,\alpha_s(\mu)} \otimes h_{j/H,b}(x;\mu) \nonumber \\
&{} \qquad + \order{\frac{\Lambda^2_{\rm QCD}}{k_c^2}}\, ,
\label{e.deltasivers}
\end{align}
analogously to the unpolarized case in \eref{yuk}, but
now allowing for mixing with lower dimensional operators. 
Similar to~\eref{yuk}, if \eref{basic} is valid up to perturbative corrections, then the collinear matrix elements on the right side of \eref{deltasivers} must be operators with 
equal or lower dimension to $T_{i(g)/H}(x;\mu)$, and the $\mathcal{C}_{ij}$ must begin at order $\alpha_s$ or higher 
and involve only the logarithms $L$ (\eref{logdef}). 
The ``??'' is to emphasize that \eref{deltasivers} is provisional and will actually turn out not to hold. 

\section{Non-Verification} 
\label{s.direct}

\begin{figure*}
\centering 
 \begin{tabular}{c@{\hspace*{5mm}}c}
  \includegraphics[width=0.4\textwidth]{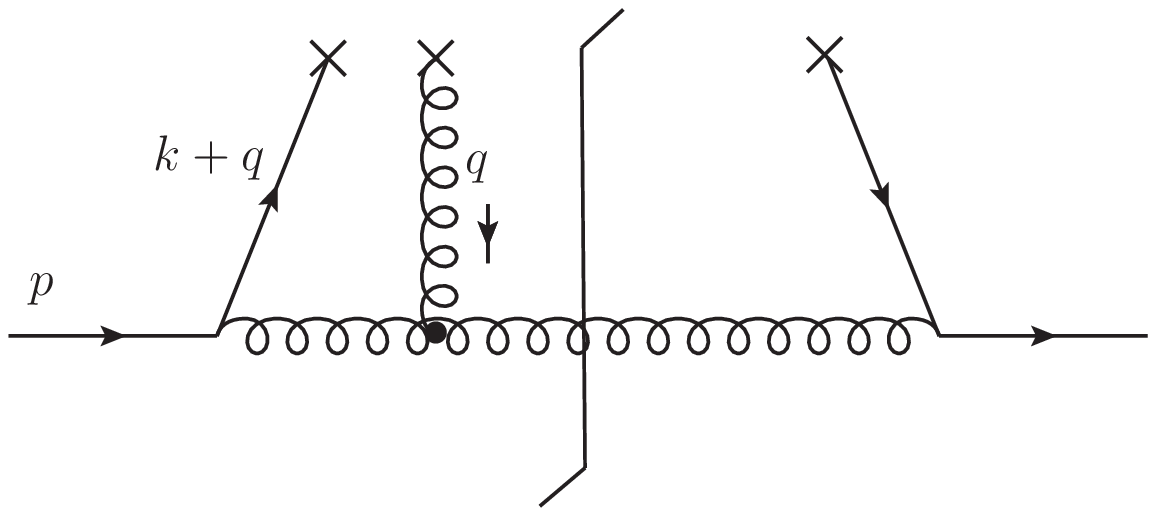}
  \hspace{0.45cm}
  &
  \hspace{0.45cm}
  \includegraphics[width=0.4\textwidth]{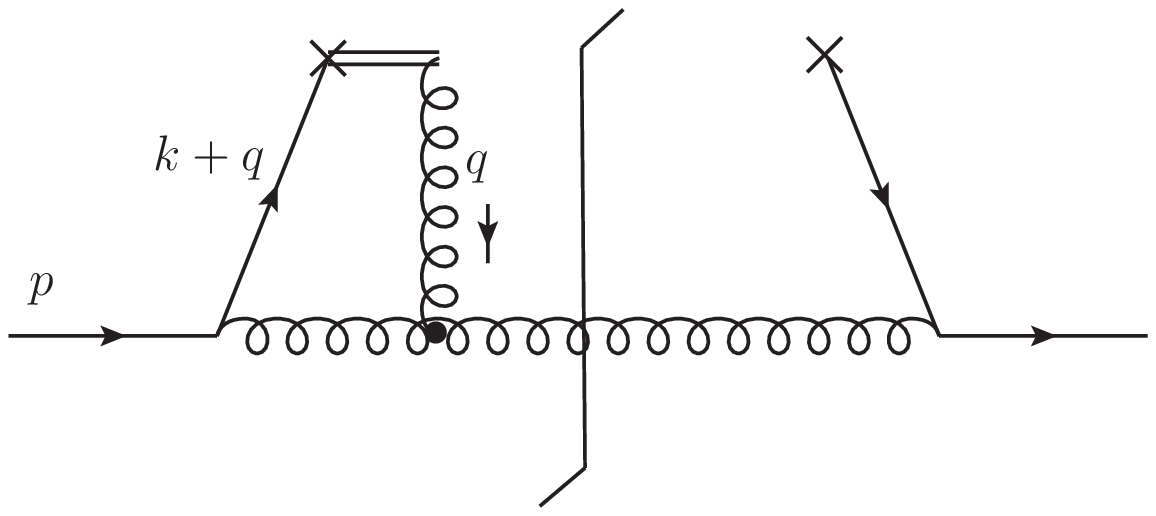} \\
  (a) & (b) 
\end{tabular}
\caption{(a) Lowest order pole part of the twist-3 quark-gluon correlation function. 
(b) Lowest order TMD Sivers function. 
The calculations are nearly identical up to the overall $-1/2M$, the factor of $k_T^2$, 
and the integral over transverse momentum in the case of the twist-3 quark-gluon correlation function.}
\label{f.feyngraph}
\end{figure*}

A complication with checking relations like 
\eref{deltasivers} directly is that the functions involved are nonperturbative. However, the   
generic behavior that we are interested in testing can be checked order-by-order in any theory with the relevant properties of renormalizability and a gauge interaction, for example in a model field theory with a spectator. 
Moreover, if the factorization in \eref{deltasivers} were true generally, then it must hold order-by-order for partonic targets. We consider, therefore,
a non-zero mass quark in pQCD as the target to remain as close to true QCD as possible,
while the quark mass $m_q$ also serves as a regulator for perturbative collinear singularities.
The lowest order non-vanishing graphs are shown in \fref{feyngraph}, 
with \fref{feyngraph}(a) representing the twist-3 collinear calculation (the second term on the right 
side of \eref{deltdef2}) and \fref{feyngraph}(b) representing the TMD pdf calculation (the integrand for the first term on the right side of \eref{deltdef2}).~\footnote{We have labeled the exchanged gluon momentum by $q$ to be consistent with the notation in \cite{Kang:2010hg}. This should not be confused, however, with a virtual photon momentum like the one 
in Fig.~\ref{fig:ssa4dy}.} 
Although we are calculating in perturbation theory, the calculation must be viewed as a kind of model
since the functions are non-perturbative, and we must assume that a suitable infrared regulator has been imposed on higher order graphs, though we will not need to make the specific regulator explicit here because all the graphs in \fref{feyngraph} are infrared 
and collinear finite with a nonzero quark mass
and a fixed momentum fraction $x$.

Both calculations in \fref{feyngraph} proceed similarly, up to the factor of $k_T^2/M^2$ and the absent integral over transverse momentum for the TMD PDF case. 
Fortunately, several 
features of \fref{feyngraph}(a,b) simplify their calculation. First, the TMD PDF case is finite in both the UV and IR, even with a zero mass gluon.
Second, if we restrict to the class of non-singular covariant gauges, they are gauge-independent as can be seen from the fact they (and their Hermitian conjugates) are the only graphs that contribute at $\order{\alpha_s}$ to the transverse single spin asymmetry with unpolarized active quark, so no subtleties associated with the Wilson line in the twist-3 quark-gluon correlation function arise. In general, other graphs are needed for gauge invariance -- see the discussion of \fref{feyngraphtu} in \aref{appendix} for more on this. Finally, the graphs contain no light-cone divergences, so subtleties associated with Wilson lines and light-cone regulators do not affect our calculations. (Of course, in more general higher order graphs, all these issues will become important.) The result is a kind 
of spectator model that closely mirrors 
actual pQCD calculations. 

Most of the steps needed to calculate each of the two terms on the right side of \eref{deltdef2} can be found in already existing literature~\cite{Kang:2010hg,Bacchetta:2008af}, with only slight modifications 
needed in the twist-3 quark-gluon correlation function case to convert to dimensional regularization and minimal subtraction.
(See~\aref{appendix} for a discussion of these calculations.) Model calculations of transverse spin and momentum effects were also calculated earlier in \cite{Ji:2002aa,Gamberg:2003ey,Goeke:2006ef,Gamberg:2007wm}.
While the complete result for $\Delta f_{1T}^\perp(k_c)$ is not relevant to our discussion, a very important result is that it involves double logarithmic  terms with the quark mass $m_q$ of the form
\begin{align}
&{} 
\Delta f_{1T}^\perp(k_c \sim \mu) \no &{} \; = 
 -\frac{C_F N_c}{{2}\pi}  \alpha_s(\mu)^2 x(1-x) \ln^2\parz{\frac{\mu}{(1-x) m_q}} + \cdots \, .
\label{e.dlog}
\end{align}
The ``$\cdots$'' refers to all other terms not involving double logarithms of the form $\ln^2(\mu/\text{mass})$.

To see that this creates complications, consider \eref{deltasivers} expanded through the first several orders,
\begin{align}
M &{} \Delta f_{1T}^\perp(k_c) = \no
& \sum_{ij} \mathcal{C}^{(2)}_{ij}\parz{x/x',L,\alpha_s(\mu)} \otimes T_{j(g)/H}^{(0)}(x;\mu) \no
& + \sum_{ij} \mathcal{C}^{(1)}_{ij}\parz{x/x',L,\alpha_s(\mu)} \otimes T_{j(g)/H}^{(1)}(x;\mu)  \no 
& + \sum_{ij,b} m_{b,ij} \mathcal{C}^{(2)}_{b,ij}\parz{x/x',L,\alpha_s(k_c)} \otimes   h^{(0)}_{b,j/H}(x;\mu) \no
& + \sum_{ij,b} m_{b,ij} \mathcal{C}^{(1)}_{b,ij}\parz{x/x',L,\alpha_s(k_c)} \otimes h^{(1)}_{b,j/H}(x;\mu)  \no
& + \text{h.o.}  + \order{\frac{\Lambda^2_{\rm QCD}}{k_c^2}} \, ,
\label{e.deltasivers2}
\end{align}
with the $(...)$ superscripts denoting the order in perturbation theory.
If \eref{deltasivers} were true, then one of these terms must contain the double logarithm in \eref{dlog}. But
\begin{equation}
T_{j(g)/H}^{(1),\perp}(x;\mu) =  T_{j(g)/H}^{(0),\perp}(x;\mu) = 0 \, ,
\end{equation}
because at least two gluons (a spectator and a final state interaction) are needed for the correlation function to be nonzero. 
So if \eref{deltasivers2} could accommodate \eref{dlog}, then the $\alpha_s^2 \ln^2(\mu/((1-x)m_q))$ 
would have to appear in either the fourth or fifth lines. However, the order-$\alpha_s^0$ 
$h^{(0)}_{b,i/H}(x;\mu)$ and the order-$\alpha_s^1$ 
$h^{(1)}_{b,i/H}(x;\mu)$ can contain at most zero and one $\ln (\mu)$ factors  respectively. This means at least one power 
of $\ln \parz{\mu/((1-x)m_q)}$ would have to be included 
inside $\mathcal{C}^{(2)}_{b,ij}$ or $\mathcal{C}^{(1)}_{b,ij}$. If this were done, however, it would violate 
the requirement that no logarithms other than the mass-independent \eref{logdef} appear in the hard $\mathcal{C}$-coefficients. 
This shows that the factorization in \eref{deltasivers2}, and therefore \eref{deltasivers} generally, is invalid. 

An equivalent and more direct way to state the above is simply to note that since the coupling only vanishes like $\alpha_s(\mu) \sim 1/\ln(\mu)$ for $\mu \gg \Lambda_\text{QCD}$, then the term in \eref{dlog} undergoes no suppression at large $\mu$. 

It should be understood that, since the correlation functions are strictly speaking nonperturbative, the mass scales like the $m_q$ in 
\eref{dlog} represent more general non-perturbative structures. 
In some ways, therefore, a model renormalizable diquark spectator theory is more illustrative of the 
problem described above, since mass scales like the $m_q$ in 
\eref{dlog} become more complicated non-perturbative objects. 

It is possibly tempting to argue that in a proton target terms like \eref{dlog} would be suppressed by $m_q/M_p$ ratios. But this same ratio appears in 
all terms to all orders in the correlation function, so there is no relative suppression. This is especially, clear in other model theories like a 
spectator diquark theory -- see \erefs{qsintLL}{sintLL}. Thus, the double logarithm in \eref{dlog} represents a kind of strong ultraviolet ambiguity that did not arise in the unpolarized case. 

Furthermore, the fact that the double logarithm in \eref{dlog} goes to infinity as the collinear regulator is removed,  $m_q\to 0$, signals that the two sides of \eref{basic} have different collinear sensitivities (as $k_T\to 0$) manifested by the divergent $k_T$-integration starting from its UV perturbative region and using dimensional regularization. The need to account for this divergent $m_q \to 0$ behavior will reappear in the treatment of the very large transverse momentum ($Q_T \sim Q$) region of physical processes like the Drell-Yan example in \sref{intro}.

Like in any QCD factorization approach to a physical observable, perturbative calculations of short-distance hard parts beyond the lowest order tree-level require perturbatively calculated and regularized partonic versions of the long-distance correlation functions to remove all soft and collinear divergences in the hard partonic scattering.  Since the moment of the Sivers TMD function and the twist-3 correlation function in the two sides of \eref{basic} have different collinear sensitivities, the use of the long-distance correlation functions for QCD factorization treatment of weighted SSAs requires caution and needs to be made consistent with a factorization formalism.   

\section{Discussion}
\label{s.discussion}

The contribution to $\Delta f_{1T}^\perp(k_c)$ in \eref{dlog} makes clear that there is 
very strong sensitivity to choices in how the ultraviolet contribution to the integral over transverse momentum for weighted-integrated asymmetries is implemented. The two 
schemes we considered were: 1.) standard collinear renormalization for the twist-3 collinear correlation function and 2.) direct integration of the TMD function (the Sivers function) with suppression of the large transverse momentum contribution. It is the latter method, however, that is almost always used in practical phenomenological applications. That is, parametrizations of the twist-3 collinear correlation function are usually constructed from the Sivers function via \eref{basic}.  

This implies that it is the evolution of the Sivers function, performed using standard TMD evolution techniques and before the integration over $k_T$, that governs the evolution of the weighted-integrated asymmetries as they are normally determined.

The technical reason for the term in \eref{dlog}
is that the box-loop integral in \fref{feyngraph}(a) produces  
a power of $(\mu/((1-x) m_q)^{2 \epsilon}$ in dimensional regularization in addition to the $(\mu/((1-x) m_q)^{2 \epsilon}$ that already comes from 
the divergent $k_T$ integral. 
In a calculation of the renormalized twist-3 function, both multiply a $1/\epsilon^2$ from the divergent $k_T$-integral to produce two 
$\ln^2(\mu/((1-x)m_q))$ terms. 
By contrast, the TMD pdf calculation is finite at the order of graphs in \fref{feyngraph}, so 
$\order{\epsilon}$ factors never contribute. 
The only relevant  
$\ln(\mu/((1-x) m_q)$ in the integral of the TMD PDF comes directly from the cut-off transverse momentum integral when it is applied
on the left side of \eref{basic}. 
The result is that the $\ln^2(\mu/((1-x)m_q))$ term in the renormalized twist-3 correlation
function comes with an extra factor of 2 compared with  
the $\ln^2(\mu/((1-x)m_q))$ term in the weighted integral of the Sivers function. Thus, the double logarithms like 
\eref{dlog} do not cancel in \eref{deltdef2}. 

The more general reason is that transverse momentum integrals do not commute with the removal of ultraviolet regulators, 
a property that has already been remarked upon in some detail in, for example, Ref.~\cite{Kang:2010hg}. 
This results in a type of scale anomaly that already appears in the unpolarized leading-twist case, \eref{closure}.
In \eref{yuk}, however, large contributions analogous to \eref{dlog} do not arise because the 
transverse momentum integrals corresponding to the particular graphs in \fref{feyngraph} are finite for the transverse momentum integral 
in \eref{closure}. 

Some physical intuition for the mismatch is gained by recalling that the design region for the TMD PDF treatment, where the approximations that give TMD factorization apply, is the very 
small $Q_T \ll Q$ region, including $Q_T \lesssim \Lambda_\text{QCD}$, whereas the behavior of the TMD PDF at $k_T$ close to physical hard scales is not physically meaningful without some correction term. But the factor of $k_T^2/M^2$ in the integrand of \eref{basic} effectively discards the relevant $k_T \sim 0$ contribution to the cross section while amplifying the ill-defined contribution
from $k_T > \mu$.  Therefore, the resulting integral is dominated by an arbitrary scheme used to regulate the large $k_T$ behavior. In other 
words, TMD factorization derivations apply to cross sections differential in $Q_T$ and in the small $Q_T$ limit, but the $Q_T$ weighting suppresses this small $Q_T$ region (in fact creating a zero) while magnifying the 
$Q_T \sim Q$ region of the cross section where a different sort of factorization is needed. That the single ${\bf Q}_T$-weighting is the lowest power ${\bf Q}_T$ weight that gives a non-zero integrated transverse SSA does not mitigate the potential for such shifts in the important momentum range to spoil relations like \eref{basic}.

The particular order in which transverse momentum integrals are evaluated and ultraviolet regulators are removed is important.
In renormalized collinear correlation functions (like the twist-3 quark-gluon correlation function), the ultraviolet regulator needs to be 
the same for real and virtual emissions for ensuring such features as the automatic cancellations of light-cone divergences in collinear correlation functions~\cite{Collins:2003fm}. Thus, 
ultraviolet regulators can only be removed after all integrals are evaluated. By contrast, in the unintegrated TMD PDFs there are no regulators on real parton transverse momentum since the transverse momentum is fixed to values determined by the physical cross section.
It is only at later stages that 
a $k_T^2$-weighted integral of a phenomenologically extracted Sivers function is performed, as in \eref{basic}, at which point a cutoff on the physical region of $k_T \gtrsim \mu$ is restored in a separate step. This reversal in the natural order of regulator removal between the two cases is the origin of the problem discussed in the previous section.

Forcing a version of \eref{basic} amounts to dealing with 
issues such as light-cone divergences in the twist-3 quark-gluon correlation function point-by-point in parton transverse momentum first, before transverse momentum integrals with real emissions are evaluated.  This allows separate ultraviolet regulators to be applied to real and virtual ultraviolet divergences. 
Then it is possible to impose the requirement that the weighted Sivers and twist-3 calculations use the same ultraviolet regulators on real emissions from the outset, thus ensuring \eref{basic}. This is equivalent to defining the TMD PDF first, and then defining the corresponding twist-3 function via the weighted transverse momentum integral of the TMD function. In this view, \eref{basic} should be viewed as a definition rather than a derived result. 
Nevertheless, such a convention preserves the logical structure embodied in 
relations like \eref{basic}, and thereby allows twist-3 calculations an interpretation in terms of intrinsic transverse momentum.\footnote{Note 
that results like \cite{Gamberg:2017jha} amount only to one of potentially many arbitrary regulator schemes for the 
integral on the left side of \eref{basic}, and are not 
actual derivations of \eref{basic}. Specifically, they do not address the question of regulator sensitivity.} 
This then provides one answer to the question of which type of scale evolution is relevant in weighted integrals  of spin asymmetries, in cases where large transverse momentum is strongly suppressed. If, as we suggest above, the collinear $T_{q(g)/H}(x)$ on the right side of \eref{basic} is defined via 
the TMD pdf on the left side, then evolution is dictated by the TMD evolution of $f_{1T,q/H}^{\perp}(x,\Tsc{k}{})$ at small transverse momentum. Of course, at 
very large $Q$ the integral becomes dominated by non-intrinsic perturbatively generated transverse momentum radiation~\cite{Grewal:2020hoc}, and a switch to a scheme like~\cite{Gamberg:2017jha} may then be useful to exploit refactorization. 

Obtaining a fully fixed prescription for treating divergences in parton correlation functions requires complete factorization treatments 
for specific processes, to clarify how those parton correlation functions 
contribute to the evaluation of corresponding hard parts. 
We emphasize that more work in this direction is needed.

A potential complication is that if the twist-3 function is defined via TMD PDFs, then it might inherit some of the problems with TMD factorization that can arise in  
hadron-hadron collisions with measured hadron transverse momentum in the final state~\cite{Collins:2007nk,Rogers:2010dm}. Such effects may be mitigated, however, if scales are evolved high enough that the integrand is dominated by a perturbatively generated tail. Moreover, a full treatment of the matching to the large $Q_T \sim Q$ region is needed. We leave the investigation of all
such issues to future work. 

\appendix

\section{Calculation of $\Delta f_{1T}^\perp(k_c)$}
\label{a.appendix}

Here we explain some of the details leading to \eref{dlog}. 
Since the 
basic integrals have all been set up before~\cite{Kang:2010hg}, we will simply refer to earlier literature, only modifying those parts needed to implement renormalization with 
dimensional regularization and minimal subtraction.
\subsection{The scalar field spectator}
\label{a.scalar}

It will be simplest to structure the argument by starting with the result for the scalar 
diquark model and explain the steps to transform to QCD.
We start from Eq.~(29) in Ref.~\cite{Kang:2010hg}\footnote{Equation~(44) of Ref.~\cite{Kang:2010hg} differs by a sign from \eref{qs} due to a different convention for the direction of the Wilson line. There it is chosen to be consistent with Drell-Yan-like processes} in the scalar model. Here we adopt a sign convention consistent with \eref{qs}, and compute the integrals over $\T{k}$ and $\T{q}$ in $n=2-2\epsilon$ dimensions. In dimensional regularization, the only $n$ dependence is from the integration measure and the factor $\mu^{4\epsilon}$ that comes with the couplings. With the point-like coupling between the nucleon, the quark, and the spectator diquark, the integral in that equation becomes, up to overall factors, 
\begin{align}
I_T\equiv 
    & \int \frac{\diff{^n\T{k}}{}}{(2\pi)^n}\frac{\diff{^n\T{q}}{}}{(2\pi)^n}
    \frac{\Tscsq{q}{}-(\T{q}\cdot \T{S})^2}{\Tscsq{q}{}[\Tscsq{k}{}+\Lambda_s^2][(\T{k}+\T{q})^2+\Lambda_s^2]} \no
    = & \frac{\pi^{n/2}\Gamma(2-n/2)}{\Gamma(2)}\int_0^1\diff{\alpha}{} \int \frac{\diff{^n\T{q}}{}}{(2\pi)^{2n}} 
    \frac{1}{\Tscsq{q}{}} \no
    & \times [\Tscsq{q}{}-(\T{q}\cdot \T{S})^2][\alpha(1-\alpha)\Tscsq{q}{}+\Lambda_s^2]^{n/2-2} \, , 
\label{e.qsint1}
\end{align}
where 
\begin{equation}
    \Lambda_s^2=xM_s^2+(1-x)m_q^2-x(1-x)M^2.
\end{equation}
with $M_s$, $m_q$, and $M$ being the masses of the scalar diquark, quark and nucleon, respectively. 
By choosing the orientations
\begin{align}
    \T{q}=&\Tsc{q}{}(\sin{\theta_1}\sin{\theta_2}\cdots\sin{\theta_{n-1}},\cdots, \cos{\theta_1}) \label{e.ndimqT} \, , \\
    \T{S}=&(0,\cdots, 1)
    \label{e.ndimST}
\end{align}
in $n$ dimensions (\eref{ndimqT} takes the form of the standard $n$ dimensional spherical coordinates, while in~\eref{ndimST} all components of $\T{S}$ are 0 except for the last one), it is straightforward to carry out the angular part of the integral and verify
\begin{equation}
    \int \diff{^n\T{q}}{}\Tscsq{q}{} = n\int \diff{^n\T{q}}{}(\T{q}\cdot \T{S})^2.
\end{equation}
Then it is valid in~\eref{qsint1} to replace $\Tscsq{q}{}-(\T{q}\cdot \T{S})^2\rightarrow (1-1/n)\Tscsq{q}{}$ and obtain
\begin{align}
I_T
    = & \frac{\pi^{n/2}\Gamma(2-n/2)}{\Gamma(2)}\parz{1-\frac{1}{n}} \no
    & \times \int_0^1\diff{\alpha}{} \int \frac{\diff{^n\T{q}}{}}{(2\pi)^n} 
    [\alpha(1-\alpha)\Tscsq{q}{}+\Lambda_s^2]^{n/2-2}. \no
    = & \frac{\pi^n}{(2\pi)^{2n}}\left(1-\frac{1}{n} \right)\Gamma(2-n)\Lambda_s^{2(n-2)} \times \no & \qquad \times \int_0^1\diff{\alpha}{}[\alpha(1-\alpha)]^{-n/2} \no
    = & \frac{\pi^n}{(2\pi)^{2n}}\left(1-\frac{1}{n}\right)\Gamma(2-n)\Lambda_s^{2(n-2)}\frac{\Gamma^2(\epsilon)}{\Gamma(2\epsilon)} \, .
\label{e.qsint2}
\end{align}
 Restoring the overall factors dropped in \eref{qsint1} and expanding near $\epsilon = 0$ gives 
the logarithmic terms with
\begin{align}
-\frac{1}{M} T_{q(g)/H}(x;\mu)
& = \frac{N_cC_Fg\lambda_s^2g_s}{16\pi^3}(1-x)\parz{\frac{m_q}{M}+x} 
     \no
& {\hskip -0.2in}\times
\Bigl( \ln^2{(\Lambda_s/\mu)} 
+ 
\frac{1}{2} \left( 1+2\gamma_E-2\ln{(4\pi)} \right)
\no
& {\hskip 0.6in} 
\times \ln{(\Lambda_s/\mu)} \Bigr) 
 + \cdots 
    \, .
\label{e.qsintLL}
\end{align}

The analog of \eref{qsint2} for the weighted Sivers function comes from the integral in Eq.~(49) in Ref.~\cite{Kang:2010hg},  which was 
also calculated in~\cite{Bacchetta:2008af}. Without overall factors the integral is
\begin{align}
    I_{S}
    \equiv & \frac{1}{4\pi M^2} \int_0^{k_c} \frac{\diff{\Tsc{k}{}}{}}{2\pi} \frac{\Tsc{k}{}}{\Tscsq{k}{}+\Lambda_s^2}\ln{\frac{\Tscsq{k}{}+\Lambda_s^2}{\Lambda_s^2}} \no
    = & \frac{1}{32\pi^2 M^2}\ln^2(k_c^2/\Lambda_s^2+1) \, .
\label{e.sint1}
\end{align}
Cutting off the $\Tsc{k}{}$ integral at $k_c = \mu$, restoring the overall factors
dropped in \eref{sint1}, and expanding to lowest order in $\Lambda_s/\mu$:
\begin{align}
&{} \int \diff{^2\T{k}}{} \frac{\Tscsq{k}{}}{M^2} f_{1T}^{\perp}(x,\Tsc{k}{}) = \nonumber \\ &{}\qquad \frac{N_cC_Fg\lambda_s^2g_s}{32\pi^3}(1-x)\parz{\frac{m_q}{M}+x}\ln^2{(\Lambda_s/\mu)} \nonumber \\
&{} \qquad + \order{\frac{\Lambda_\text{QCD}^2}{\mu^2} } \, .
\label{e.sintLL}
\end{align}
Note the factor of two difference between the double logarithmic terms in \eref{sintLL} and \eref{qsintLL}. 
Subtracting \eref{sintLL} and \eref{qsintLL} gives a version 
of \eref{dlog} for the case of a scalar field for the spectator. 

\begin{figure}
\centering 
 \begin{tabular}{c}
  \includegraphics[width=0.4\textwidth]{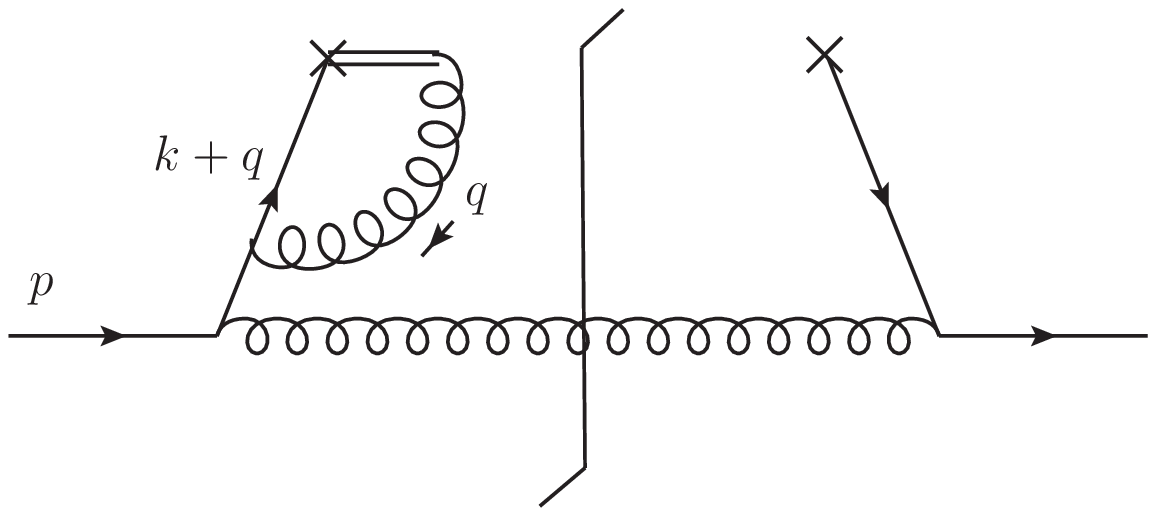} \\
  (a) \\
  \vspace{.2in} \\
  \includegraphics[width=0.4\textwidth]{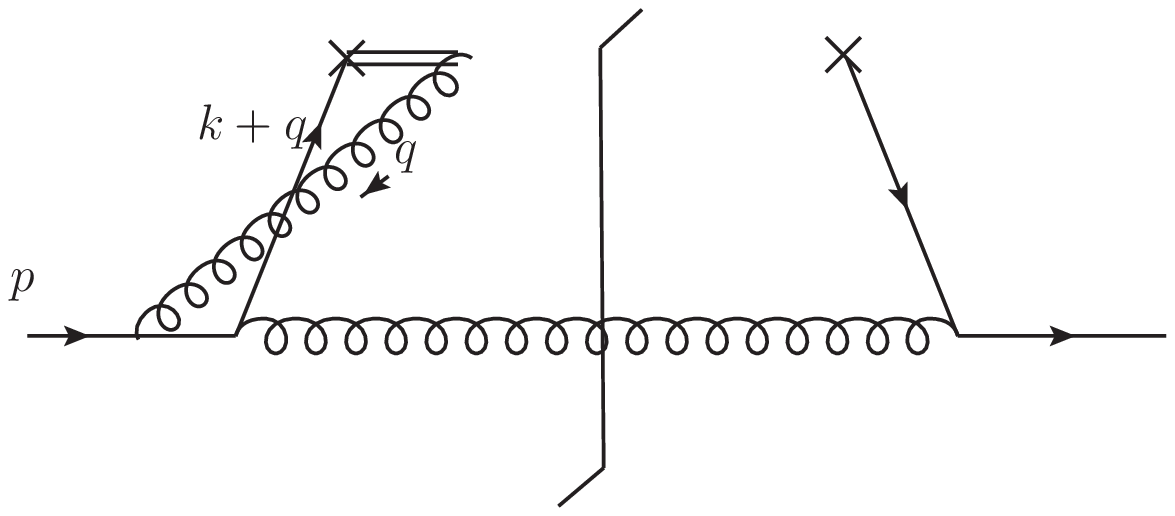} \\
  (b) \\
  \vspace{.1in} 
\end{tabular}
\caption{Graphs (a) and (b), along with their Hermitian conjugates, are needed in general 
for gauge invariance. Analogous graphs are also needed for the collinear twist-three function.}
\label{f.feyngraphtu}
\end{figure}

\subsection{QCD}
Transitioning to the case of QCD with an incoming quark of mass $m_q$ primarily entails a change in the Dirac trace. 
The trace part of the Sivers calculation
in the diquark model is
\begin{align}
	{\rm Tr}_\text{S}^\text{diquark}\equiv & {\rm Tr}\left[\gamma^+(\sla{k}+\sla{q}+m_q)(\sla{p}+M)\gamma^5 \sla{s_T}(\sla{k}+m_q)\right] \no
	& \times (2p-2k-q)^{\tau}n_{\tau} \no
	= & 8i{p^+}^2(1-x)(Mx+m_q)\epsilon^{ij}q_{Ti}s_{Tj} \, ,
\label{e.trSdiqk}
\end{align}
where the $i \pi \delta(q^+)$ from the eikonal propagator of the Wilson line constrains $q^+ = 0$ and gives the imaginary part of the $q$ integral
(see Eq.(94) of Ref.~\cite{Bacchetta:2008af}). In QCD this trace becomes
\begin{align}
	{\rm Tr}_\text{S}^\text{QCD}\equiv &{\rm Tr}\left[\gamma^+(\sla{k}+\sla{q}+m_q)\gamma^{\mu}(\sla{p}+m_q)\gamma^5 \sla{s_T}\gamma^{\alpha}(\sla{k}+m_q)\right] \no
	& d_{\mu\nu}(p-k-q) 
	d_{\sigma\alpha}(p-k) 
	 \times \bigl[(2p-2k-q)^{\tau}g^{\nu\sigma} \no  &+(k-p-q)^{\nu}g^{\sigma\tau}+(2q-p+k)^{\sigma}g^{\tau\nu} \bigr]n_{\tau} \no
	=& 16i{p^+}^2xm_q(x-1)\epsilon^{ij}q_{Ti}s_{Tj} \, , 
\label{e.trSqcd}
\end{align}
where we work in Feynman gauge with the gluon polarization tensor (no ghost graphs contribute at this order):
\begin{equation}
	d_{\mu\nu}(q)= -g_{\mu\nu} \, .
\end{equation}
Similarly, the traces for the twist-3 quark-gluon correlation function in the two theories are
\begin{align}
	{\rm Tr}_\text{T}^\text{diquark} &\equiv  {\rm Tr}\left[\gamma^+(\sla{k}+\sla{q}+m_q)(\sla{p}+M)\gamma^5 \sla{s_T}(\sla{k}+m_q)\right] \no
	& \epsilon^{s_T\rho n \bar{n}}(2p-2k-q)^{\tau}(-n_{\tau}q_{\rho}) \no
	=& -8i{p^+}^2(1-x)(Mx+m_q) \no
	& \times [\Tscsq{q}{}-(\T{q}\cdot \T{S})^2]\, , \\
	{\rm Tr}_\text{T}^\text{QCD} & \equiv \no & {\rm Tr}\left[\gamma^+(\sla{k}+\sla{q}+m_q)\gamma^{\mu}(\sla{p}+m_q)\gamma^5 \sla{s_T}\gamma^{\alpha}(\sla{k}+m_q)\right] \no
	& d_{\mu\nu}(p-k-q)
	d_{\sigma\alpha}(p-k) \no
	& \times \bigl[(2p-2k-q)^{\tau}g^{\nu\sigma} 
	+(k-p-q)^{\nu}g^{\sigma\tau} \no
	& +(2q-p+k)^{\sigma}g^{\tau\nu} \bigr]\epsilon^{s_T\rho n \bar{n}}(-n_{\tau}q_{\rho}) \no
	=&-8i{p^+}^2(nx+2-n)m_q(x-1) \no
	& \times [\Tscsq{q}{}-(\T{q}\cdot \T{S})^2] \, .
\end{align}
The trace part of the twist-3 quark-gluon correlation function in QCD acquires a dependence on the dimension $n$. Note that for Sivers function we always work in 4 dimensions ($n=2$), so ${\rm Tr}_\text{S}^\text{QCD}$ does not have a similar factor. 
In the case of twist-3 correlation function, $q^+=0$ is imposed by the delta function from the cut vertex~\cite{Kang:2008ey,Kang:2010hg}. The $n$ and 
$\bar{n}$ four-vectors in the above traces represent light-like vectors in the minus and plus directions respectively, not to be confused with the spacetime dimension.

In general, the graphs in \fref{feyngraphtu} are also needed to give gauge invariance, but we have confirmed 
that they do not contribute in either the twist 3 collinear or TMD calculations when the target is transversely polarized and the active quark is unpolarized. 

The momentum integrals in QCD and diquark model are  almost identical for both Sivers and collinear twist-3 functions, with only the replacement
\begin{equation}
    \Lambda_s^2\rightarrow \Lambda_g^2\equiv (1-x)^2m_q^2.
\end{equation}
Also note that in the transition to QCD the change of the coupling constants and the color factors are the same for Sivers and twist-3 PDFs,
\begin{equation}
    gg_s\lambda_s^2\rightarrow 16\pi^2\alpha_S^2, \quad N_cC_F\rightarrow -\frac{1}{2}N_cC_F \, .
\end{equation}
Aside from the above replacements, for Sivers in QCD one needs $(1-x)(Mx+m_q)\rightarrow 2m_qx(x-1)$, as can be seen from~\eref{trSdiqk} and \eref{trSqcd}. The resulting logarithmic terms are
\begin{equation}
    L_S = \frac{N_cC_F\alpha_S^2}{2\pi}x(1-x)\ln^2{(\Lambda_g/\mu)} \, . 
\end{equation}
For the twist-3 quark-gluon correlation function however, we must change $(1-x)(Mx+m_q)\rightarrow m_q(nx+2-n)(x-1)$ and include the factor $(nx+2-n)$ when expanding near $\epsilon=0$ if we wish to obtain all logarithms beyond the double logarithm. The result is
\begin{align}
    L_T = &-\frac{N_cC_F\alpha_S^2}{\pi}x(1-x)\biggl (\ln^2{(\Lambda_g/\mu)} \no
    &+\left( 1+\gamma_E-\ln{(4\pi)}-\frac{1}{2x} \right) \ln{(\Lambda_g/\mu)}\biggr ) \, .
\end{align}
From~\eref{deltdef2}
\begin{align}
\Delta f_{1T}^\perp(k_c)=&L_T+L_S+\cdots \no
=&-\frac{N_cC_F\alpha_S^2}{2\pi}x(1-x)\ln^2{(\Lambda_g/\mu)}+\cdots \, .
\end{align}

\vskip 0.3in
\acknowledgments
We thank 
F.~Aslan, 
L.~Gamberg, A. Metz and D.~Pitonyak for helpful discussion.  This work is supported in part by the U.S. Department of Energy contract DE-AC05-06OR23177, under which Jefferson Science Associates, LLC, manages and operates Jefferson Lab.
T.~Rogers was also supported by the U.S. Department of Energy, Office of Science, Office of Nuclear Physics, under Award Number DE-SC0018106.
B.~Wang is supported in part by the National Science Foundation of China (11875232) and the Zhejiang University Fundamental Research Funds for the Central Universities.
This work is also supported in part within the framework of the TMD Topical Collaboration. 
\bibliography{bibliography}

\providecommand{\noopsort}[1]{}
\begin{thebibliography}{70}
\expandafter\ifx\csname natexlab\endcsname\relax\def\natexlab#1{#1}\fi
\expandafter\ifx\csname bibnamefont\endcsname\relax
  \def\bibnamefont#1{#1}\fi
\expandafter\ifx\csname bibfnamefont\endcsname\relax
  \def\bibfnamefont#1{#1}\fi
\expandafter\ifx\csname citenamefont\endcsname\relax
  \def\citenamefont#1{#1}\fi
\expandafter\ifx\csname url\endcsname\relax
  \def\url#1{\texttt{#1}}\fi
\expandafter\ifx\csname urlprefix\endcsname\relax\def\urlprefix{URL }\fi
\providecommand{\bibinfo}[2]{#2}
\providecommand{\eprint}[2][]{\url{#2}}

\bibitem[{\citenamefont{Bunce et~al.}(1976)}]{Bunce:1976yb}
\bibinfo{author}{\bibfnamefont{G.}~\bibnamefont{Bunce}} \bibnamefont{et~al.},
  \bibinfo{journal}{Phys. Rev. Lett.} \textbf{\bibinfo{volume}{36}},
  \bibinfo{pages}{1113} (\bibinfo{year}{1976}).

\bibitem[{\citenamefont{Adams et~al.}(1991)}]{Adams:1991rw}
\bibinfo{author}{\bibfnamefont{D.~L.} \bibnamefont{Adams}} \bibnamefont{et~al.}
  (\bibinfo{collaboration}{E581, E704}), \bibinfo{journal}{Phys. Lett.}
  \textbf{\bibinfo{volume}{B261}}, \bibinfo{pages}{201} (\bibinfo{year}{1991}).

\bibitem[{\citenamefont{Bravar et~al.}(1996)}]{Bravar:1996ki}
\bibinfo{author}{\bibfnamefont{A.}~\bibnamefont{Bravar}} \bibnamefont{et~al.}
  (\bibinfo{collaboration}{Fermilab E704}), \bibinfo{journal}{Phys. Rev. Lett.}
  \textbf{\bibinfo{volume}{77}}, \bibinfo{pages}{2626} (\bibinfo{year}{1996}).

\bibitem[{\citenamefont{Aidala et~al.}(2012)\citenamefont{Aidala, Bass, Hasch,
  and Mallot}}]{Aidala:2012mv}
\bibinfo{author}{\bibfnamefont{C.~A.} \bibnamefont{Aidala}},
  \bibinfo{author}{\bibfnamefont{S.~D.} \bibnamefont{Bass}},
  \bibinfo{author}{\bibfnamefont{D.}~\bibnamefont{Hasch}}, \bibnamefont{and}
  \bibinfo{author}{\bibfnamefont{G.~K.} \bibnamefont{Mallot}}
  (\bibinfo{year}{2012}), \eprint{1209.2803}.

\bibitem[{\citenamefont{Kane et~al.}(1978)\citenamefont{Kane, Pumplin, and
  Repko}}]{Kane:1978nd}
\bibinfo{author}{\bibfnamefont{G.~L.} \bibnamefont{Kane}},
  \bibinfo{author}{\bibfnamefont{J.}~\bibnamefont{Pumplin}}, \bibnamefont{and}
  \bibinfo{author}{\bibfnamefont{W.}~\bibnamefont{Repko}},
  \bibinfo{journal}{Phys. Rev. Lett.} \textbf{\bibinfo{volume}{41}},
  \bibinfo{pages}{1689} (\bibinfo{year}{1978}).

\bibitem[{\citenamefont{Accardi et~al.}(2016)}]{Accardi:2012qut}
\bibinfo{author}{\bibfnamefont{A.}~\bibnamefont{Accardi}} \bibnamefont{et~al.},
  \bibinfo{journal}{Eur. Phys. J.} \textbf{\bibinfo{volume}{A52}},
  \bibinfo{pages}{268} (\bibinfo{year}{2016}), \eprint{1212.1701}.

\bibitem[{\citenamefont{Collins et~al.}(1989)\citenamefont{Collins, Soper, and
  Sterman}}]{Collins:1989gx}
\bibinfo{author}{\bibfnamefont{J.~C.} \bibnamefont{Collins}},
  \bibinfo{author}{\bibfnamefont{D.~E.} \bibnamefont{Soper}}, \bibnamefont{and}
  \bibinfo{author}{\bibfnamefont{G.~F.} \bibnamefont{Sterman}},
  \bibinfo{journal}{Adv. Ser. Direct. High Energy Phys.}
  \textbf{\bibinfo{volume}{5}}, \bibinfo{pages}{1} (\bibinfo{year}{1989}),
  \eprint{hep-ph/0409313}.

\bibitem[{\citenamefont{Sivers}(1990)}]{Sivers:1989cc}
\bibinfo{author}{\bibfnamefont{D.~W.} \bibnamefont{Sivers}},
  \bibinfo{journal}{Phys. Rev.} \textbf{\bibinfo{volume}{D41}},
  \bibinfo{pages}{83} (\bibinfo{year}{1990}).

\bibitem[{\citenamefont{Sivers}(1991)}]{Sivers:1990fh}
\bibinfo{author}{\bibfnamefont{D.~W.} \bibnamefont{Sivers}},
  \bibinfo{journal}{Phys. Rev.} \textbf{\bibinfo{volume}{D43}},
  \bibinfo{pages}{261} (\bibinfo{year}{1991}).

\bibitem[{\citenamefont{Brodsky et~al.}(2002)\citenamefont{Brodsky, Hwang, and
  Schmidt}}]{Brodsky:2002rv}
\bibinfo{author}{\bibfnamefont{S.~J.} \bibnamefont{Brodsky}},
  \bibinfo{author}{\bibfnamefont{D.~S.} \bibnamefont{Hwang}}, \bibnamefont{and}
  \bibinfo{author}{\bibfnamefont{I.}~\bibnamefont{Schmidt}},
  \bibinfo{journal}{Nucl. Phys.} \textbf{\bibinfo{volume}{B642}},
  \bibinfo{pages}{344} (\bibinfo{year}{2002}), \eprint{hep-ph/0206259}.

\bibitem[{\citenamefont{Efremov and Teryaev}(1985)}]{Efremov:1984ip}
\bibinfo{author}{\bibfnamefont{A.~V.} \bibnamefont{Efremov}} \bibnamefont{and}
  \bibinfo{author}{\bibfnamefont{O.~V.} \bibnamefont{Teryaev}},
  \bibinfo{journal}{Phys. Lett.} \textbf{\bibinfo{volume}{150B}},
  \bibinfo{pages}{383} (\bibinfo{year}{1985}).

\bibitem[{\citenamefont{Qiu and Sterman}(1991{\natexlab{a}})}]{Qiu:1991pp}
\bibinfo{author}{\bibfnamefont{J.-W.} \bibnamefont{Qiu}} \bibnamefont{and}
  \bibinfo{author}{\bibfnamefont{G.~F.} \bibnamefont{Sterman}},
  \bibinfo{journal}{Phys. Rev. Lett.} \textbf{\bibinfo{volume}{67}},
  \bibinfo{pages}{2264} (\bibinfo{year}{1991}{\natexlab{a}}).

\bibitem[{\citenamefont{Qiu and Sterman}(1992)}]{Qiu:1991wg}
\bibinfo{author}{\bibfnamefont{J.-W.} \bibnamefont{Qiu}} \bibnamefont{and}
  \bibinfo{author}{\bibfnamefont{G.~F.} \bibnamefont{Sterman}},
  \bibinfo{journal}{Nucl. Phys.} \textbf{\bibinfo{volume}{B378}},
  \bibinfo{pages}{52} (\bibinfo{year}{1992}).

\bibitem[{\citenamefont{Qiu and Sterman}(1999)}]{Qiu:1998ia}
\bibinfo{author}{\bibfnamefont{J.-W.} \bibnamefont{Qiu}} \bibnamefont{and}
  \bibinfo{author}{\bibfnamefont{G.~F.} \bibnamefont{Sterman}},
  \bibinfo{journal}{Phys. Rev.} \textbf{\bibinfo{volume}{D59}},
  \bibinfo{pages}{014004} (\bibinfo{year}{1999}), \eprint{hep-ph/9806356}.

\bibitem[{\citenamefont{Collins et~al.}(1985)\citenamefont{Collins, Soper, and
  Sterman}}]{Collins:1984kg}
\bibinfo{author}{\bibfnamefont{J.~C.} \bibnamefont{Collins}},
  \bibinfo{author}{\bibfnamefont{D.~E.} \bibnamefont{Soper}}, \bibnamefont{and}
  \bibinfo{author}{\bibfnamefont{G.}~\bibnamefont{Sterman}},
  \bibinfo{journal}{Nucl. Phys.} \textbf{\bibinfo{volume}{B250}},
  \bibinfo{pages}{199} (\bibinfo{year}{1985}).

\bibitem[{\citenamefont{Collins}(2011)}]{Collins:2011zzd}
\bibinfo{author}{\bibfnamefont{J.}~\bibnamefont{Collins}},
  \bibinfo{journal}{Camb. Monogr. Part. Phys. Nucl. Phys. Cosmol.}
  \textbf{\bibinfo{volume}{32}}, \bibinfo{pages}{1} (\bibinfo{year}{2011}).

\bibitem[{\citenamefont{Ji et~al.}(2005)\citenamefont{Ji, Ma, and
  Yuan}}]{Ji:2004wu}
\bibinfo{author}{\bibfnamefont{X.-D.} \bibnamefont{Ji}},
  \bibinfo{author}{\bibfnamefont{J.-P.} \bibnamefont{Ma}}, \bibnamefont{and}
  \bibinfo{author}{\bibfnamefont{F.}~\bibnamefont{Yuan}},
  \bibinfo{journal}{Phys. Rev.} \textbf{\bibinfo{volume}{D71}},
  \bibinfo{pages}{034005} (\bibinfo{year}{2005}), \eprint{hep-ph/0404183}.

\bibitem[{\citenamefont{Ji et~al.}(2004)\citenamefont{Ji, Ma, and
  Yuan}}]{Ji:2004xq}
\bibinfo{author}{\bibfnamefont{X.-D.} \bibnamefont{Ji}},
  \bibinfo{author}{\bibfnamefont{J.-P.} \bibnamefont{Ma}}, \bibnamefont{and}
  \bibinfo{author}{\bibfnamefont{F.}~\bibnamefont{Yuan}},
  \bibinfo{journal}{Phys. Lett.} \textbf{\bibinfo{volume}{B597}},
  \bibinfo{pages}{299} (\bibinfo{year}{2004}), \eprint{hep-ph/0405085}.

\bibitem[{\citenamefont{Qiu and Sterman}(1991{\natexlab{b}})}]{Qiu:1990xy}
\bibinfo{author}{\bibfnamefont{J.-W.} \bibnamefont{Qiu}} \bibnamefont{and}
  \bibinfo{author}{\bibfnamefont{G.}~\bibnamefont{Sterman}},
  \bibinfo{journal}{Nucl. Phys.} \textbf{\bibinfo{volume}{B353}},
  \bibinfo{pages}{137} (\bibinfo{year}{1991}{\natexlab{b}}).

\bibitem[{\citenamefont{Qiu and Sterman}(1991{\natexlab{c}})}]{Qiu:1990cu}
\bibinfo{author}{\bibfnamefont{J.-W.} \bibnamefont{Qiu}} \bibnamefont{and}
  \bibinfo{author}{\bibfnamefont{G.~F.} \bibnamefont{Sterman}},
  \bibinfo{journal}{AIP Conf. Proc.} \textbf{\bibinfo{volume}{223}},
  \bibinfo{pages}{249} (\bibinfo{year}{1991}{\natexlab{c}}).

\bibitem[{\citenamefont{Ji et~al.}(2006{\natexlab{a}})\citenamefont{Ji, Qiu,
  Vogelsang, and Yuan}}]{Ji:2006ub}
\bibinfo{author}{\bibfnamefont{X.}~\bibnamefont{Ji}},
  \bibinfo{author}{\bibfnamefont{J.-W.} \bibnamefont{Qiu}},
  \bibinfo{author}{\bibfnamefont{W.}~\bibnamefont{Vogelsang}},
  \bibnamefont{and} \bibinfo{author}{\bibfnamefont{F.}~\bibnamefont{Yuan}},
  \bibinfo{journal}{Phys. Rev. Lett.} \textbf{\bibinfo{volume}{97}},
  \bibinfo{pages}{082002} (\bibinfo{year}{2006}{\natexlab{a}}),
  \eprint{hep-ph/0602239}.

\bibitem[{\citenamefont{Ji et~al.}(2006{\natexlab{b}})\citenamefont{Ji, Qiu,
  Vogelsang, and Yuan}}]{Ji:2006vf}
\bibinfo{author}{\bibfnamefont{X.}~\bibnamefont{Ji}},
  \bibinfo{author}{\bibfnamefont{J.-W.} \bibnamefont{Qiu}},
  \bibinfo{author}{\bibfnamefont{W.}~\bibnamefont{Vogelsang}},
  \bibnamefont{and} \bibinfo{author}{\bibfnamefont{F.}~\bibnamefont{Yuan}},
  \bibinfo{journal}{Phys. Rev.} \textbf{\bibinfo{volume}{D73}},
  \bibinfo{pages}{094017} (\bibinfo{year}{2006}{\natexlab{b}}),
  \eprint{hep-ph/0604023}.

\bibitem[{\citenamefont{Scimemi et~al.}(2019)\citenamefont{Scimemi, Tarasov,
  and Vladimirov}}]{Scimemi:2019gge}
\bibinfo{author}{\bibfnamefont{I.}~\bibnamefont{Scimemi}},
  \bibinfo{author}{\bibfnamefont{A.}~\bibnamefont{Tarasov}}, \bibnamefont{and}
  \bibinfo{author}{\bibfnamefont{A.}~\bibnamefont{Vladimirov}},
  \bibinfo{journal}{JHEP} \textbf{\bibinfo{volume}{05}}, \bibinfo{pages}{125}
  (\bibinfo{year}{2019}), \eprint{1901.04519}.

\bibitem[{\citenamefont{Kang et~al.}(2013)\citenamefont{Kang, Vitev, and
  Xing}}]{Kang:2012ns}
\bibinfo{author}{\bibfnamefont{Z.-B.} \bibnamefont{Kang}},
  \bibinfo{author}{\bibfnamefont{I.}~\bibnamefont{Vitev}}, \bibnamefont{and}
  \bibinfo{author}{\bibfnamefont{H.}~\bibnamefont{Xing}},
  \bibinfo{journal}{Phys. Rev.} \textbf{\bibinfo{volume}{D87}},
  \bibinfo{pages}{034024} (\bibinfo{year}{2013}), \eprint{1212.1221}.

\bibitem[{\citenamefont{Dai et~al.}(2015)\citenamefont{Dai, Kang, Prokudin, and
  Vitev}}]{Dai:2014ala}
\bibinfo{author}{\bibfnamefont{L.-Y.} \bibnamefont{Dai}},
  \bibinfo{author}{\bibfnamefont{Z.-B.} \bibnamefont{Kang}},
  \bibinfo{author}{\bibfnamefont{A.}~\bibnamefont{Prokudin}}, \bibnamefont{and}
  \bibinfo{author}{\bibfnamefont{I.}~\bibnamefont{Vitev}},
  \bibinfo{journal}{Phys. Rev.} \textbf{\bibinfo{volume}{D92}},
  \bibinfo{pages}{114024} (\bibinfo{year}{2015}), \eprint{1409.5851}.

\bibitem[{\citenamefont{Gamberg et~al.}(2018)\citenamefont{Gamberg, Metz,
  Pitonyak, and Prokudin}}]{Gamberg:2017jha}
\bibinfo{author}{\bibfnamefont{L.}~\bibnamefont{Gamberg}},
  \bibinfo{author}{\bibfnamefont{A.}~\bibnamefont{Metz}},
  \bibinfo{author}{\bibfnamefont{D.}~\bibnamefont{Pitonyak}}, \bibnamefont{and}
  \bibinfo{author}{\bibfnamefont{A.}~\bibnamefont{Prokudin}},
  \bibinfo{journal}{Phys. Lett.} \textbf{\bibinfo{volume}{B781}},
  \bibinfo{pages}{443} (\bibinfo{year}{2018}), \eprint{1712.08116}.

\bibitem[{\citenamefont{Xing and Yoshida}(2019)}]{Xing:2019nof}
\bibinfo{author}{\bibfnamefont{H.}~\bibnamefont{Xing}} \bibnamefont{and}
  \bibinfo{author}{\bibfnamefont{S.}~\bibnamefont{Yoshida}},
  \bibinfo{journal}{Adv. High Energy Phys.} \textbf{\bibinfo{volume}{2019}},
  \bibinfo{pages}{4825790} (\bibinfo{year}{2019}), \eprint{1904.00416}.

\bibitem[{\citenamefont{Luo and Sun}(2020)}]{Luo:2020hki}
\bibinfo{author}{\bibfnamefont{X.}~\bibnamefont{Luo}} \bibnamefont{and}
  \bibinfo{author}{\bibfnamefont{H.}~\bibnamefont{Sun}},
  \bibinfo{journal}{Phys. Rev. D} \textbf{\bibinfo{volume}{101}},
  \bibinfo{pages}{074016} (\bibinfo{year}{2020}), \eprint{2004.03764}.

\bibitem[{\citenamefont{Berger and Qiu}(2003)}]{Berger:2003pd}
\bibinfo{author}{\bibfnamefont{E.~L.} \bibnamefont{Berger}} \bibnamefont{and}
  \bibinfo{author}{\bibfnamefont{J.-W.} \bibnamefont{Qiu}},
  \bibinfo{journal}{Phys.\ Rev.\ Lett.} \textbf{\bibinfo{volume}{91}},
  \bibinfo{pages}{222003} (\bibinfo{year}{2003}), \eprint{hep-ph/0304267}.

\bibitem[{\citenamefont{Boer et~al.}(2003)\citenamefont{Boer, Mulders, and
  Pijlman}}]{Boer:2003cm}
\bibinfo{author}{\bibfnamefont{D.}~\bibnamefont{Boer}},
  \bibinfo{author}{\bibfnamefont{P.~J.} \bibnamefont{Mulders}},
  \bibnamefont{and} \bibinfo{author}{\bibfnamefont{F.}~\bibnamefont{Pijlman}},
  \bibinfo{journal}{Nucl. Phys.} \textbf{\bibinfo{volume}{B667}},
  \bibinfo{pages}{201} (\bibinfo{year}{2003}), \eprint{hep-ph/0303034}.

\bibitem[{\citenamefont{Kang et~al.}(2011)\citenamefont{Kang, Qiu, Vogelsang,
  and Yuan}}]{Kang:2011hk}
\bibinfo{author}{\bibfnamefont{Z.-B.} \bibnamefont{Kang}},
  \bibinfo{author}{\bibfnamefont{J.-W.} \bibnamefont{Qiu}},
  \bibinfo{author}{\bibfnamefont{W.}~\bibnamefont{Vogelsang}},
  \bibnamefont{and} \bibinfo{author}{\bibfnamefont{F.}~\bibnamefont{Yuan}},
  \bibinfo{journal}{Phys. Rev.} \textbf{\bibinfo{volume}{D83}},
  \bibinfo{pages}{094001} (\bibinfo{year}{2011}), \eprint{1103.1591}.

\bibitem[{\citenamefont{Kang and Prokudin}(2012)}]{Kang:2012xf}
\bibinfo{author}{\bibfnamefont{Z.-B.} \bibnamefont{Kang}} \bibnamefont{and}
  \bibinfo{author}{\bibfnamefont{A.}~\bibnamefont{Prokudin}},
  \bibinfo{journal}{Phys. Rev.} \textbf{\bibinfo{volume}{D85}},
  \bibinfo{pages}{074008} (\bibinfo{year}{2012}), \eprint{1201.5427}.

\bibitem[{\citenamefont{Metz et~al.}(2015)\citenamefont{Metz, Pitonyak,
  Schäfer, Schlegel, Vogelsang, and Zhou}}]{Metz:2014bba}
\bibinfo{author}{\bibfnamefont{A.}~\bibnamefont{Metz}},
  \bibinfo{author}{\bibfnamefont{D.}~\bibnamefont{Pitonyak}},
  \bibinfo{author}{\bibfnamefont{A.}~\bibnamefont{Schäfer}},
  \bibinfo{author}{\bibfnamefont{M.}~\bibnamefont{Schlegel}},
  \bibinfo{author}{\bibfnamefont{W.}~\bibnamefont{Vogelsang}},
  \bibnamefont{and} \bibinfo{author}{\bibfnamefont{J.}~\bibnamefont{Zhou}},
  \bibinfo{journal}{Few Body Syst.} \textbf{\bibinfo{volume}{56}},
  \bibinfo{pages}{331} (\bibinfo{year}{2015}).

\bibitem[{\citenamefont{Gamberg et~al.}(2017)\citenamefont{Gamberg, Kang,
  Pitonyak, and Prokudin}}]{Gamberg:2017gle}
\bibinfo{author}{\bibfnamefont{L.}~\bibnamefont{Gamberg}},
  \bibinfo{author}{\bibfnamefont{Z.-B.} \bibnamefont{Kang}},
  \bibinfo{author}{\bibfnamefont{D.}~\bibnamefont{Pitonyak}}, \bibnamefont{and}
  \bibinfo{author}{\bibfnamefont{A.}~\bibnamefont{Prokudin}},
  \bibinfo{journal}{Phys. Lett.} \textbf{\bibinfo{volume}{B770}},
  \bibinfo{pages}{242} (\bibinfo{year}{2017}), \eprint{1701.09170}.

\bibitem[{\citenamefont{Kanazawa et~al.}(2014)\citenamefont{Kanazawa, Koike,
  Metz, and Pitonyak}}]{Kanazawa:2014dca}
\bibinfo{author}{\bibfnamefont{K.}~\bibnamefont{Kanazawa}},
  \bibinfo{author}{\bibfnamefont{Y.}~\bibnamefont{Koike}},
  \bibinfo{author}{\bibfnamefont{A.}~\bibnamefont{Metz}}, \bibnamefont{and}
  \bibinfo{author}{\bibfnamefont{D.}~\bibnamefont{Pitonyak}},
  \bibinfo{journal}{Phys. Rev.} \textbf{\bibinfo{volume}{D89}},
  \bibinfo{pages}{111501} (\bibinfo{year}{2014}), \eprint{1404.1033}.

\bibitem[{\citenamefont{Gamberg and Kang}(2011)}]{Gamberg:2010tj}
\bibinfo{author}{\bibfnamefont{L.}~\bibnamefont{Gamberg}} \bibnamefont{and}
  \bibinfo{author}{\bibfnamefont{Z.-B.} \bibnamefont{Kang}},
  \bibinfo{journal}{Phys. Lett.} \textbf{\bibinfo{volume}{B696}},
  \bibinfo{pages}{109} (\bibinfo{year}{2011}), \eprint{1009.1936}.

\bibitem[{\citenamefont{Alexeev et~al.}(2019)}]{Alexeev:2018zvl}
\bibinfo{author}{\bibfnamefont{M.~G.} \bibnamefont{Alexeev}}
  \bibnamefont{et~al.} (\bibinfo{collaboration}{COMPASS}),
  \bibinfo{journal}{Nucl. Phys.} \textbf{\bibinfo{volume}{B940}},
  \bibinfo{pages}{34} (\bibinfo{year}{2019}), \eprint{1809.02936}.

\bibitem[{\citenamefont{Kang and Qiu}(2009)}]{Kang:2008ey}
\bibinfo{author}{\bibfnamefont{Z.-B.} \bibnamefont{Kang}} \bibnamefont{and}
  \bibinfo{author}{\bibfnamefont{J.-W.} \bibnamefont{Qiu}},
  \bibinfo{journal}{Phys. Rev.} \textbf{\bibinfo{volume}{D79}},
  \bibinfo{pages}{016003} (\bibinfo{year}{2009}), \eprint{0811.3101}.

\bibitem[{\citenamefont{Braun et~al.}(2009)\citenamefont{Braun, Manashov, and
  Pirnay}}]{Braun:2009mi}
\bibinfo{author}{\bibfnamefont{V.}~\bibnamefont{Braun}},
  \bibinfo{author}{\bibfnamefont{A.}~\bibnamefont{Manashov}}, \bibnamefont{and}
  \bibinfo{author}{\bibfnamefont{B.}~\bibnamefont{Pirnay}},
  \bibinfo{journal}{Phys.\ Rev.\ D} \textbf{\bibinfo{volume}{80}},
  \bibinfo{pages}{114002} (\bibinfo{year}{2009}), \bibinfo{note}{[Erratum:
  Phys.Rev.D 86, 119902 (2012)]}, \eprint{0909.3410}.

\bibitem[{\citenamefont{Collins and
  Soper}(1982{\natexlab{a}})}]{Collins:1981va}
\bibinfo{author}{\bibfnamefont{J.~C.} \bibnamefont{Collins}} \bibnamefont{and}
  \bibinfo{author}{\bibfnamefont{D.~E.} \bibnamefont{Soper}},
  \bibinfo{journal}{Nucl. Phys.} \textbf{\bibinfo{volume}{B197}},
  \bibinfo{pages}{446} (\bibinfo{year}{1982}{\natexlab{a}}).

\bibitem[{\citenamefont{Aybat et~al.}(2012)\citenamefont{Aybat, Collins, Qiu,
  and Rogers}}]{Aybat:2011ge}
\bibinfo{author}{\bibfnamefont{S.~M.} \bibnamefont{Aybat}},
  \bibinfo{author}{\bibfnamefont{J.~C.} \bibnamefont{Collins}},
  \bibinfo{author}{\bibfnamefont{J.-W.} \bibnamefont{Qiu}}, \bibnamefont{and}
  \bibinfo{author}{\bibfnamefont{T.~C.} \bibnamefont{Rogers}},
  \bibinfo{journal}{Phys. Rev.} \textbf{\bibinfo{volume}{D85}},
  \bibinfo{pages}{034043} (\bibinfo{year}{2012}), \eprint{1110.6428}.

\bibitem[{\citenamefont{Kang et~al.}(2010)\citenamefont{Kang, Qiu, and
  Zhang}}]{Kang:2010hg}
\bibinfo{author}{\bibfnamefont{Z.-B.} \bibnamefont{Kang}},
  \bibinfo{author}{\bibfnamefont{J.-W.} \bibnamefont{Qiu}}, \bibnamefont{and}
  \bibinfo{author}{\bibfnamefont{H.}~\bibnamefont{Zhang}},
  \bibinfo{journal}{Phys. Rev.} \textbf{\bibinfo{volume}{D81}},
  \bibinfo{pages}{114030} (\bibinfo{year}{2010}), \eprint{1004.4183}.

\bibitem[{\citenamefont{Anselmino et~al.}(2012)\citenamefont{Anselmino,
  Boglione, and Melis}}]{Anselmino:2012aa}
\bibinfo{author}{\bibfnamefont{M.}~\bibnamefont{Anselmino}},
  \bibinfo{author}{\bibfnamefont{M.}~\bibnamefont{Boglione}}, \bibnamefont{and}
  \bibinfo{author}{\bibfnamefont{S.}~\bibnamefont{Melis}},
  \bibinfo{journal}{Phys. Rev.} \textbf{\bibinfo{volume}{D86}},
  \bibinfo{pages}{014028} (\bibinfo{year}{2012}), \eprint{1204.1239}.

\bibitem[{\citenamefont{Boglione et~al.}(2018)\citenamefont{Boglione, D'Alesio,
  Flore, and Gonzalez-Hernandez}}]{Boglione:2018dqd}
\bibinfo{author}{\bibfnamefont{M.}~\bibnamefont{Boglione}},
  \bibinfo{author}{\bibfnamefont{U.}~\bibnamefont{D'Alesio}},
  \bibinfo{author}{\bibfnamefont{C.}~\bibnamefont{Flore}}, \bibnamefont{and}
  \bibinfo{author}{\bibfnamefont{J.~O.} \bibnamefont{Gonzalez-Hernandez}},
  \bibinfo{journal}{JHEP} \textbf{\bibinfo{volume}{07}}, \bibinfo{pages}{148}
  (\bibinfo{year}{2018}), \eprint{1806.10645}.

\bibitem[{\citenamefont{Cammarota et~al.}(2020)\citenamefont{Cammarota,
  Gamberg, Kang, Miller, Pitonyak, Prokudin, Rogers, and
  Sato}}]{Cammarota:2020qcw}
\bibinfo{author}{\bibfnamefont{J.}~\bibnamefont{Cammarota}},
  \bibinfo{author}{\bibfnamefont{L.}~\bibnamefont{Gamberg}},
  \bibinfo{author}{\bibfnamefont{Z.-B.} \bibnamefont{Kang}},
  \bibinfo{author}{\bibfnamefont{J.~A.} \bibnamefont{Miller}},
  \bibinfo{author}{\bibfnamefont{D.}~\bibnamefont{Pitonyak}},
  \bibinfo{author}{\bibfnamefont{A.}~\bibnamefont{Prokudin}},
  \bibinfo{author}{\bibfnamefont{T.~C.} \bibnamefont{Rogers}},
  \bibnamefont{and} \bibinfo{author}{\bibfnamefont{N.}~\bibnamefont{Sato}}
  (\bibinfo{year}{2020}), \eprint{2002.08384}.

\bibitem[{\citenamefont{Metz and Pitonyak}(2013)}]{Metz:2012ct}
\bibinfo{author}{\bibfnamefont{A.}~\bibnamefont{Metz}} \bibnamefont{and}
  \bibinfo{author}{\bibfnamefont{D.}~\bibnamefont{Pitonyak}},
  \bibinfo{journal}{Phys. Lett.} \textbf{\bibinfo{volume}{B723}},
  \bibinfo{pages}{365} (\bibinfo{year}{2013}), \bibinfo{note}{[Erratum: Phys.
  Lett.B762,549(2016)]}, \eprint{1212.5037}.

\bibitem[{\citenamefont{Yuan and Zhou}(2009)}]{Yuan:2009dw}
\bibinfo{author}{\bibfnamefont{F.}~\bibnamefont{Yuan}} \bibnamefont{and}
  \bibinfo{author}{\bibfnamefont{J.}~\bibnamefont{Zhou}},
  \bibinfo{journal}{Phys. Rev. Lett.} \textbf{\bibinfo{volume}{103}},
  \bibinfo{pages}{052001} (\bibinfo{year}{2009}), \eprint{0903.4680}.

\bibitem[{\citenamefont{Kang et~al.}(2015)\citenamefont{Kang, Prokudin, Sun,
  and Yuan}}]{Kang:2015msa}
\bibinfo{author}{\bibfnamefont{Z.-B.} \bibnamefont{Kang}},
  \bibinfo{author}{\bibfnamefont{A.}~\bibnamefont{Prokudin}},
  \bibinfo{author}{\bibfnamefont{P.}~\bibnamefont{Sun}}, \bibnamefont{and}
  \bibinfo{author}{\bibfnamefont{F.}~\bibnamefont{Yuan}}
  (\bibinfo{year}{2015}), \eprint{1505.05589}.

\bibitem[{\citenamefont{Mulders and Tangerman}(1996)}]{Mulders:1995dh}
\bibinfo{author}{\bibfnamefont{P.~J.} \bibnamefont{Mulders}} \bibnamefont{and}
  \bibinfo{author}{\bibfnamefont{R.~D.} \bibnamefont{Tangerman}},
  \bibinfo{journal}{Nucl. Phys.} \textbf{\bibinfo{volume}{B461}},
  \bibinfo{pages}{197} (\bibinfo{year}{1996}), \bibinfo{note}{[Erratum: Nucl.
  Phys.B484,538(1997)]}, \eprint{hep-ph/9510301}.

\bibitem[{\citenamefont{Collins and
  Soper}(1982{\natexlab{b}})}]{Collins:1981uw}
\bibinfo{author}{\bibfnamefont{J.~C.} \bibnamefont{Collins}} \bibnamefont{and}
  \bibinfo{author}{\bibfnamefont{D.~E.} \bibnamefont{Soper}},
  \bibinfo{journal}{Nucl. Phys.} \textbf{\bibinfo{volume}{B194}},
  \bibinfo{pages}{445} (\bibinfo{year}{1982}{\natexlab{b}}).

\bibitem[{\citenamefont{Diehl}(2016)}]{Diehl:2015uka}
\bibinfo{author}{\bibfnamefont{M.}~\bibnamefont{Diehl}}, \bibinfo{journal}{Eur.
  Phys. J.} \textbf{\bibinfo{volume}{A52}}, \bibinfo{pages}{149}
  (\bibinfo{year}{2016}), \eprint{1512.01328}.

\bibitem[{\citenamefont{Rogers}(2016)}]{Rogers:2015sqa}
\bibinfo{author}{\bibfnamefont{T.~C.} \bibnamefont{Rogers}},
  \bibinfo{journal}{Eur. Phys. J.} \textbf{\bibinfo{volume}{A52}},
  \bibinfo{pages}{153} (\bibinfo{year}{2016}), \eprint{1509.04766}.

\bibitem[{\citenamefont{Stewart et~al.}(2010)\citenamefont{Stewart, Tackmann,
  and Waalewijn}}]{Stewart:2009yx}
\bibinfo{author}{\bibfnamefont{I.~W.} \bibnamefont{Stewart}},
  \bibinfo{author}{\bibfnamefont{F.~J.} \bibnamefont{Tackmann}},
  \bibnamefont{and} \bibinfo{author}{\bibfnamefont{W.~J.}
  \bibnamefont{Waalewijn}}, \bibinfo{journal}{Phys. Rev. D}
  \textbf{\bibinfo{volume}{81}}, \bibinfo{pages}{094035}
  (\bibinfo{year}{2010}), \eprint{0910.0467}.

\bibitem[{\citenamefont{Becher and Neubert}(2011)}]{Becher:2010tm}
\bibinfo{author}{\bibfnamefont{T.}~\bibnamefont{Becher}} \bibnamefont{and}
  \bibinfo{author}{\bibfnamefont{M.}~\bibnamefont{Neubert}},
  \bibinfo{journal}{Eur. Phys. J.} \textbf{\bibinfo{volume}{C71}},
  \bibinfo{pages}{1665} (\bibinfo{year}{2011}), \eprint{1007.4005}.

\bibitem[{\citenamefont{Becher et~al.}(2012)\citenamefont{Becher, Neubert, and
  Wilhelm}}]{Becher:2011xn}
\bibinfo{author}{\bibfnamefont{T.}~\bibnamefont{Becher}},
  \bibinfo{author}{\bibfnamefont{M.}~\bibnamefont{Neubert}}, \bibnamefont{and}
  \bibinfo{author}{\bibfnamefont{D.}~\bibnamefont{Wilhelm}},
  \bibinfo{journal}{JHEP} \textbf{\bibinfo{volume}{02}}, \bibinfo{pages}{124}
  (\bibinfo{year}{2012}), \eprint{1109.6027}.

\bibitem[{\citenamefont{Becher et~al.}(2013)\citenamefont{Becher, Neubert, and
  Wilhelm}}]{Becher:2012yn}
\bibinfo{author}{\bibfnamefont{T.}~\bibnamefont{Becher}},
  \bibinfo{author}{\bibfnamefont{M.}~\bibnamefont{Neubert}}, \bibnamefont{and}
  \bibinfo{author}{\bibfnamefont{D.}~\bibnamefont{Wilhelm}},
  \bibinfo{journal}{JHEP} \textbf{\bibinfo{volume}{05}}, \bibinfo{pages}{110}
  (\bibinfo{year}{2013}), \eprint{1212.2621}.

\bibitem[{\citenamefont{Echevarr\'{\i}a
  et~al.}(2012)\citenamefont{Echevarr\'{\i}a, Idilbi, and
  Scimemi}}]{GarciaEchevarria:2011rb}
\bibinfo{author}{\bibfnamefont{M.~G.} \bibnamefont{Echevarr\'{\i}a}},
  \bibinfo{author}{\bibfnamefont{A.}~\bibnamefont{Idilbi}}, \bibnamefont{and}
  \bibinfo{author}{\bibfnamefont{I.}~\bibnamefont{Scimemi}},
  \bibinfo{journal}{JHEP} \textbf{\bibinfo{volume}{1207}}, \bibinfo{pages}{002}
  (\bibinfo{year}{2012}), \eprint{1111.4996}.

\bibitem[{\citenamefont{Echevarria et~al.}(2013)\citenamefont{Echevarria,
  Idilbi, and Scimemi}}]{Echevarria:2012js}
\bibinfo{author}{\bibfnamefont{M.~G.} \bibnamefont{Echevarria}},
  \bibinfo{author}{\bibfnamefont{A.}~\bibnamefont{Idilbi}}, \bibnamefont{and}
  \bibinfo{author}{\bibfnamefont{I.}~\bibnamefont{Scimemi}},
  \bibinfo{journal}{Phys.Lett.} \textbf{\bibinfo{volume}{B726}},
  \bibinfo{pages}{795} (\bibinfo{year}{2013}), \eprint{1211.1947}.

\bibitem[{\citenamefont{Echevarria et~al.}(2014)\citenamefont{Echevarria,
  Idilbi, and Scimemi}}]{Echevarria:2014rua}
\bibinfo{author}{\bibfnamefont{M.~G.} \bibnamefont{Echevarria}},
  \bibinfo{author}{\bibfnamefont{A.}~\bibnamefont{Idilbi}}, \bibnamefont{and}
  \bibinfo{author}{\bibfnamefont{I.}~\bibnamefont{Scimemi}},
  \bibinfo{journal}{Phys. Rev.} \textbf{\bibinfo{volume}{D90}},
  \bibinfo{pages}{014003} (\bibinfo{year}{2014}), \eprint{1402.0869}.

\bibitem[{\citenamefont{Chiu et~al.}(2012)\citenamefont{Chiu, Jain, Neill, and
  Rothstein}}]{Chiu:2012ir}
\bibinfo{author}{\bibfnamefont{J.-Y.} \bibnamefont{Chiu}},
  \bibinfo{author}{\bibfnamefont{A.}~\bibnamefont{Jain}},
  \bibinfo{author}{\bibfnamefont{D.}~\bibnamefont{Neill}}, \bibnamefont{and}
  \bibinfo{author}{\bibfnamefont{I.~Z.} \bibnamefont{Rothstein}},
  \bibinfo{journal}{JHEP} \textbf{\bibinfo{volume}{05}}, \bibinfo{pages}{084}
  (\bibinfo{year}{2012}), \eprint{1202.0814}.

\bibitem[{\citenamefont{Li et~al.}(2016)\citenamefont{Li, Neill, and
  Zhu}}]{Li:2016axz}
\bibinfo{author}{\bibfnamefont{Y.}~\bibnamefont{Li}},
  \bibinfo{author}{\bibfnamefont{D.}~\bibnamefont{Neill}}, \bibnamefont{and}
  \bibinfo{author}{\bibfnamefont{H.~X.} \bibnamefont{Zhu}}
  (\bibinfo{year}{2016}), \eprint{1604.00392}.

\bibitem[{\citenamefont{Bacchetta et~al.}(2008)\citenamefont{Bacchetta, Conti,
  and Radici}}]{Bacchetta:2008af}
\bibinfo{author}{\bibfnamefont{A.}~\bibnamefont{Bacchetta}},
  \bibinfo{author}{\bibfnamefont{F.}~\bibnamefont{Conti}}, \bibnamefont{and}
  \bibinfo{author}{\bibfnamefont{M.}~\bibnamefont{Radici}},
  \bibinfo{journal}{Phys. Rev.} \textbf{\bibinfo{volume}{D78}},
  \bibinfo{pages}{074010} (\bibinfo{year}{2008}), \eprint{0807.0323}.

\bibitem[{\citenamefont{Ji and Yuan}(2002)}]{Ji:2002aa}
\bibinfo{author}{\bibfnamefont{X.-D.} \bibnamefont{Ji}} \bibnamefont{and}
  \bibinfo{author}{\bibfnamefont{F.}~\bibnamefont{Yuan}},
  \bibinfo{journal}{Phys. Lett.} \textbf{\bibinfo{volume}{B543}},
  \bibinfo{pages}{66} (\bibinfo{year}{2002}), \eprint{hep-ph/0206057}.

\bibitem[{\citenamefont{Gamberg et~al.}(2003)\citenamefont{Gamberg, Goldstein,
  and Oganessyan}}]{Gamberg:2003ey}
\bibinfo{author}{\bibfnamefont{L.~P.} \bibnamefont{Gamberg}},
  \bibinfo{author}{\bibfnamefont{G.~R.} \bibnamefont{Goldstein}},
  \bibnamefont{and} \bibinfo{author}{\bibfnamefont{K.~A.}
  \bibnamefont{Oganessyan}}, \bibinfo{journal}{Phys. Rev.}
  \textbf{\bibinfo{volume}{D67}}, \bibinfo{pages}{071504}
  (\bibinfo{year}{2003}), \eprint{hep-ph/0301018}.

\bibitem[{\citenamefont{Goeke et~al.}(2006)\citenamefont{Goeke, Meissner, Metz,
  and Schlegel}}]{Goeke:2006ef}
\bibinfo{author}{\bibfnamefont{K.}~\bibnamefont{Goeke}},
  \bibinfo{author}{\bibfnamefont{S.}~\bibnamefont{Meissner}},
  \bibinfo{author}{\bibfnamefont{A.}~\bibnamefont{Metz}}, \bibnamefont{and}
  \bibinfo{author}{\bibfnamefont{M.}~\bibnamefont{Schlegel}},
  \bibinfo{journal}{Phys.\ Lett.\ B} \textbf{\bibinfo{volume}{637}},
  \bibinfo{pages}{241} (\bibinfo{year}{2006}), \eprint{hep-ph/0601133}.

\bibitem[{\citenamefont{Gamberg et~al.}(2008)\citenamefont{Gamberg, Goldstein,
  and Schlegel}}]{Gamberg:2007wm}
\bibinfo{author}{\bibfnamefont{L.~P.} \bibnamefont{Gamberg}},
  \bibinfo{author}{\bibfnamefont{G.~R.} \bibnamefont{Goldstein}},
  \bibnamefont{and} \bibinfo{author}{\bibfnamefont{M.}~\bibnamefont{Schlegel}},
  \bibinfo{journal}{Phys. Rev.} \textbf{\bibinfo{volume}{D77}},
  \bibinfo{pages}{094016} (\bibinfo{year}{2008}), \eprint{0708.0324}.

\bibitem[{\citenamefont{Collins}(2003)}]{Collins:2003fm}
\bibinfo{author}{\bibfnamefont{J.~C.} \bibnamefont{Collins}},
  \bibinfo{journal}{Acta Phys. Polon.} \textbf{\bibinfo{volume}{B34}},
  \bibinfo{pages}{3103} (\bibinfo{year}{2003}), \eprint{hep-ph/0304122}.

\bibitem[{\citenamefont{Grewal et~al.}(2020)\citenamefont{Grewal, Kang, Qiu,
  and Signori}}]{Grewal:2020hoc}
\bibinfo{author}{\bibfnamefont{M.}~\bibnamefont{Grewal}},
  \bibinfo{author}{\bibfnamefont{Z.-B.} \bibnamefont{Kang}},
  \bibinfo{author}{\bibfnamefont{J.-W.} \bibnamefont{Qiu}}, \bibnamefont{and}
  \bibinfo{author}{\bibfnamefont{A.}~\bibnamefont{Signori}}
  (\bibinfo{year}{2020}), \eprint{2003.07453}.

\bibitem[{\citenamefont{Collins and Qiu}(2007)}]{Collins:2007nk}
\bibinfo{author}{\bibfnamefont{J.}~\bibnamefont{Collins}} \bibnamefont{and}
  \bibinfo{author}{\bibfnamefont{J.-W.} \bibnamefont{Qiu}},
  \bibinfo{journal}{Phys. Rev.} \textbf{\bibinfo{volume}{D75}},
  \bibinfo{pages}{114014} (\bibinfo{year}{2007}), \eprint{0705.2141}.

\bibitem[{\citenamefont{Rogers and Mulders}(2010)}]{Rogers:2010dm}
\bibinfo{author}{\bibfnamefont{T.~C.} \bibnamefont{Rogers}} \bibnamefont{and}
  \bibinfo{author}{\bibfnamefont{P.~J.} \bibnamefont{Mulders}},
  \bibinfo{journal}{Phys. Rev.} \textbf{\bibinfo{volume}{D81}},
  \bibinfo{pages}{094006} (\bibinfo{year}{2010}), \eprint{1001.2977}.

\end{thebibliography}

\end{document}